\documentclass[aps,PRApplied,twocolumn,superscriptaddress,floatfix]{revtex4}
\usepackage{graphicx}
\usepackage{latexsym}
\usepackage{stmaryrd}
\usepackage{amsmath}
\usepackage{amssymb}
\usepackage{float}
\usepackage[colorlinks,linkcolor=blue,anchorcolor=blue,urlcolor=blue,citecolor=blue]{hyperref}
\begin{document}
\title{Two-dimensional metals for piezotronics devices based on Berry curvature dipole}
\author{Rui-Chun Xiao}
\affiliation{School of Physical Science and Technology, Soochow University, Suzhou, 215006, China}
\affiliation{Institute for Advanced Study, Soochow University, Suzhou 215006, China}
\author{Ding-Fu Shao}
\affiliation{Department of Physics and Astronomy Nebraska Center for Materials and Nanoscience, University of Nebraska, Lincoln, Nebraska 68588-0299, USA}
\author{Zhi-Qiang Zhang}
\affiliation{School of Physical Science and Technology, Soochow University, Suzhou, 215006, China}
\author{Hua Jiang}\email{jianghuaphy@suda.edu.cn}
\affiliation{School of Physical Science and Technology, Soochow University, Suzhou, 215006, China}
\affiliation{Institute for Advanced Study, Soochow University, Suzhou 215006, China}

\begin{abstract}
Piezotronics is an emerging field, which exploits strain to control the transport properties in condensed matters. At present, piezotronics research majorly focuses on insulators with tunable electric dipole by strain. Metals are excluded in this type of applications due to the absence of electric dipole. The recently discovered Berry curvature dipole can exist in metals, thus introduces the possibility of the piezotronics phenomena in them. In this paper, we predict that strain can switch the Berry curvature dipole, and lead the nonlinear Hall effect in the two-dimensional (2D) 1$H$-MX$_2$ (M=Nb, Ta; X=S, Se). Based on symmetry analysis and first-principles calculations, we show these 2D monolayer metals have the desired piezotronics property: without strain the Berry curvature dipole is eliminated by symmetry, prohibiting the nonlinear Hall effect; while uniaxial strain can effectively reduce the symmetry to introduce sizable Berry curvature dipole, and it can generate observable Hall voltage in a reasonable experimental condition.
Due to the nonlinear and topological properties, the piezoelectricity here is quite different from the traditional one based on the electric dipole. 
Compared with the traditional piezoelectronic materials which can only be presented in insulators, we manifest that the 2D metallic 1$H$-MX$_2$ (M=Nb, Ta; X=S, Se) are also the ideal platform for piezotronics.

\end{abstract}
\maketitle
\section{Introduction}
The piezotronics as a new research field has attracted growing interests for promising electronic applications in sensors, transducers, power generation, \emph{etc.} \cite{RN2248, RN2245, RN2235, RN2250, RN2251, RN2244, RN2233, RN2249, RN2234, RN2252}. Piezotronics exploits the piezoelectric potential created in materials to control the transport properties. Special interests have been focused on two-dimensional (2D) piezoelectric materials \cite{RN2232, RN2231, RN2342}, due to the convenience of strain application and ability to withstand the enormous strain. Electric dipole $\mathbf{P}$ can be orderly arranged by strain in solids, due to the transitions between non-polar and polar symmetries. Such a mechanism becomes the focus of piezoelectricity research \cite{RN2232, RN2337, RN2239,RN2231, RN2231, RN2240, RN2345, RN2342, RN2345}. The current piezotronics researches majorly focus on the insulators, because $\mathbf{P}$ in metals will be neutralized by free electrons. Exploration of an analogy of $\mathbf{P}$ will be helpful to lead piezo-response in metal for promising piezotronics applications.

\begin{figure}
\begin{flushleft}
\centering
\includegraphics[width=1\columnwidth]{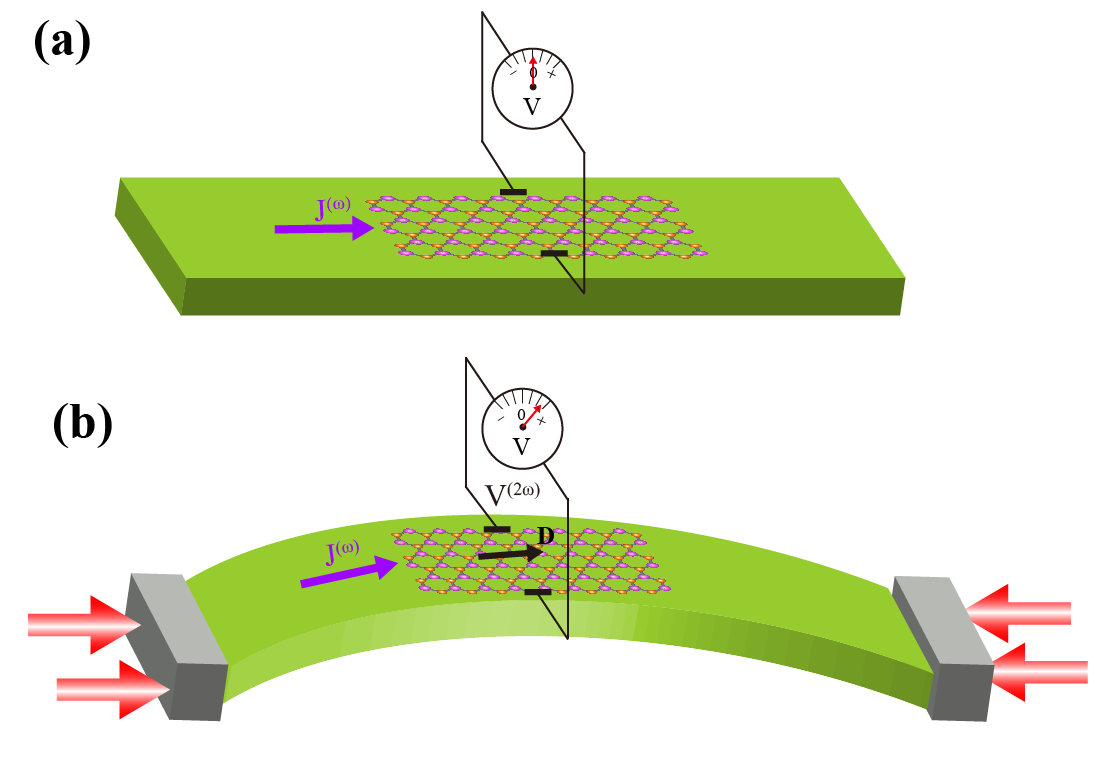}
\caption{A metallic piezotronics device based on Berry curvature dipole $\mathbf{D}$. Charge current $J^{(\omega)}$ leads to (a) zero nonlinear Hall voltage without strain due to the vanishing of $\mathbf{D}$, and (b) finite nonlinear Hall voltage $V^{(2\omega)}$ with uniaxial strain due to the emerging of $\mathbf{D}$.}
\label{Illa}
\end{flushleft}
\end{figure}

The Berry curvature dipole $\mathbf{D}$ \cite{RN2111} was proposed recently and might provide an avenue to introduce desired piezotronics properties in metals. $\mathbf{D}$ is the measure of the separation of positive and negative Berry curvature in reciprocal space. $\mathbf{D}$ relates to the electronic states at the Fermi surface, so it can only exist in metals. $\mathbf{D}$ can generate second-order anomalous transport phenomena in the absence of magnetic field, \emph{i.e.} nonlinear Hall effect \cite{RN2111}, which was confirmed in bilayer 2D WTe$_2$ by experiments \cite{RN2105, RN2110, RN2158} recently. Similar with $\mathbf{P}$ \cite{RN2345}, the emergence of $\mathbf{D}$ also requires the crystal symmetry \cite{RN2111}. Except for bilayer 2D WTe$_2$, the symmetries of most 2D materials force the $\mathbf{D}$ to be zero \cite{RN1966, RN1891}. Since strain can be easily applied in 2D materials to change the symmetry and switch $\mathbf{D}$, a 2D metal piezotronics device based on nonlinear Hall effect may be fabricated, as shown in Fig. \ref{Illa}. In such a device, $\mathbf{D}$ is zero for the pristine 2D metal, leading to zero nonlinear Hall voltage [Fig. \ref{Illa} (a)]. Strain can break the crystal symmetry, lead finite $\mathbf{D}$, and may generate observable nonlinear Hall voltage [Fig. \ref{Illa} (b)].

1$H$-phase monolayers transition-metal dichalcogenides (TMDs) are one of the most investigated 2D material categories, and they are the candidates to realize such a piezotronics device. Uniaxial strain can be used to reduce the symmetry of 1$H$-MX$_2$ and switch $\mathbf{D}$. It has been discussed that the uniaxial strain can introduce the nonlinear Hall effect in 1$H$-MoS$_2$ and 1$H$-WSe$_2$\cite{RN2111, RN2113, RN2172}. However, these two materials are semiconductors, the emerging of $\mathbf{D}$ also requires the complemental electric doping, which limits the piezotronics application based on the nonlinear Hall effect. Intrinsic 2D metallic 1$H$-MX$_2$ (M=Nb, Ta; X=S, Se) might be more promising candidates to realize such piezotronics device. The strain should be able to switch the $\mathbf{D}$ in 2D metals, as strain manipulates $\mathbf{P}$ in semiconductor 1$H$-MoS$_2$ \cite{RN2232, RN2342, RN2231,RN2345}.

In this work, using symmetry analysis and first-principles calculations, we show that monolayer metallic 1$H$-MX$_2$ (M=Nb, Ta; X=S, Se) are very promising for the piezotronics applications based on Berry curvature dipole. Without strain, $\mathbf{D}$ in these materials is zero due to the symmetry restriction, prohibiting the nonlinear Hall effect. Small uniaxial strain (along zigzag or armchair direction) can effectively reduce the symmetry and generate sizable $\mathbf{D}$, leading to observable nonlinear Hall voltage in a conventional experimental condition. Since the traditional piezoelectricity induced by electric dipole $\mathbf{P}$ has wide applications in both electronic and piezotronic devices \cite{RN2345}, the piezoelectricity based on the Berry curvature dipole $\mathbf{D}$ should also have a significant impact on the develop of piezotronics devices. 

The rest of the paper is organized as follows. In Section II, we introduce how to use the strain to break the symmetry and switch $\mathbf{D}$ in 1$H$-MX$_2$ (M=Nb, Ta; X=S, Se). Section III, we introduce the first-principles calculation methods. The calculation results and corresponding explanations are shown in Section IV. Discussion of the characters of piezoelectricity and a brief conclusion are listed in Section V.

\section{Symmetry analysis}

In order to get the ideal piezotronics effect, the nonlinear Hall effect should be very small or zero without strain, while greatly enhanced under strain. Since nonlinear Hall signal is proportional to Berry curvature dipole $\mathbf{D}$, and $\mathbf{D}$ is high related to the symmetry. We analysis the symmetries of $\mathbf{D}$ with and without strain for 1$H$-MX$_2$ (M=Nb, Ta; X=S, Se) first.

Since 1$H$-MX$_2$ (M=Nb, Ta; X=S, Se) has time-reversal symmetry $\mathcal{T}$, but no inversion symmetry $\mathcal{I}$, we start with analyzing the Berry curvature dipole $\mathbf{D}$ under these two types of symmetry.
As preliminary, we analysis the symmetry of Berry curvature ${{\Omega}_{n, a}}(\mathbf{k})$ first. ${{\Omega}_{n, a}}(\mathbf{k})$ is odd under the time-reversal symmetry, \emph{i.e.} $\mathcal{T}$: ${{\Omega}_{n,a}}(\mathbf{k})=-{{\Omega }_{n,a}}(-\mathbf{k})$, where $n$ is the band index and $a\in\{x,y,z\}$. Therefore the integration of $\Omega_{n, a}(\boldsymbol{k})$ of occupied states in a non-magnetic (time-reversal invariant) material is zero, and it is the reason why the anomalous Hall effect can not appear in non-magnetic materials. Besides, ${{\Omega}_{n, a}}(\mathbf{k})$ is even with respect to inversion symmetry, $\mathcal{I}$: ${{\Omega}_{n,a}}(\mathbf{k})={{\Omega}_{n,a}}(-\mathbf{k})$. Hence, under time-reversal symmetry $\mathcal{T}$ and inversion symmetry $\mathcal{I}$, the ${{\Omega}_{n,a}}(\mathbf{k})$ is zero at every $\mathbf{k}$ point. Therefore it is necessary to break the inversion symmetry $\mathcal{I}$ to obtain the local Berry curvature in the non-magnetic systems.

Berry curvature dipole $\mathbf{D}$ is a 3$\times$3 tensor, and ${{D}_{bd}}$ is an element of $\mathbf{D}$, defined as:
\begin{align}
D_{bd}=-\sum_{n} \int \frac{\partial f_{n}(\mathbf{k})}{\partial \epsilon_{n}(\mathbf{k})} v_{b} \Omega_{n, d}(\mathbf{k}) d[\mathbf{k}]=\int d_{b d} d[\mathbf{k}],
\label{EqDbd}
\end{align}
where ${{d}_{bd}}(\mathbf{k})$ is the Berry curvature dipole density
\begin{align}
d_{bd}(\mathbf{k})=-\sum_{n} \frac{\partial f_{n}(\mathbf{k})}{\partial \epsilon_{n}(\mathbf{k})} v_{b} \Omega_{n, d}(\mathbf{k}),
\label{dbd}
\end{align}
${{f}_{n}}(\mathbf{k})$ refers to the Fermi distribution, and $v_{b}=\frac{\partial \epsilon_{n}(\mathbf{k})}{\partial \mathbf{k}_{b}}$ is the electron velocity. The factor $\frac{\partial f_{n}(\mathbf{k})}{\partial \epsilon_{n}(\mathbf{k})}$ in Eq. (\ref{EqDbd}) implies ${{D}_{bd}}$ is a Fermi surface determined quantity. $d_{bd}(\mathbf{k})$ is even with respect to the time reversal symmetry and odd under the inversion symmetry according to Eq. (\ref{dbd}), which is opposite to the case of the Berry curvature, \emph{i.e},
\begin{align}
\left\{\begin{array}{l}
  \mathcal{T}:{{d}_{bd}}(\mathbf{k})={{d}_{bd}}(-\mathbf{k}), \\
  \mathcal{I}:{{d}_{bd}}(\mathbf{k})=-{{d}_{bd}}(-\mathbf{k}). \\
\end{array} \right.
\label{Eq_dbd}
\end{align}
Therefore, when $\mathcal{I}$ is broken, nonzero ${{D}_{bd}}$ can emerge in a non-magnetic metal, \emph{i.e.} the nonlinear Hall current might appear.

Different from $\mathbf{P}$, $\mathbf{D}$ is a pseudo-tensor and determined by \cite{RN2111}
\begin{align}
	\mathbf{D}=\det (S)S\mathbf{D}{{S}^{-1}},
\label{Dsymm}
\end{align}
where $S$ denotes the symmetric operation matrix of the point group. Specifically, since Berry curvature ${{\Omega }_{n, a}}(\mathbf{k})$ only has a $z$ component in a 2D system, only ${{D}_{xz}}$ and ${{D}_{yz}}$ exist in 2D materials. In other words, $\mathbf{D}$ transforms as a pseudovector in the 2D systems.
In 1$H$ monolayers transition-metal dichalcogenides, the point group $D_{3h}$ ($\bar{6}m2$) contains a three-fold rotation $\mathcal{C}_{3}$ symmetry along $z$ direction ($\mathcal{C}_{3z}$), one mirror reflection perpendicular to $z$ direction ($\mathcal{M}_{xy}$), three mirrors are parallel to $z$-direction and three in-plane $\mathcal{C}_{2}$ symmetries [Fig. \ref{bandstru}(a)]. The $\mathcal{C}_{3z}$, $\mathcal{C}_{2y}$ and $\mathcal{M}_{xy}$ are the generator operators of ${{D}_{3h}}$:
\begin{align}
   \mathcal{C}_{3z}=\left( \begin{matrix}
   \cos ({2\pi }/{3}\;) & -\sin ({2\pi }/{3}\;) & 0  \\
   \sin ({2\pi }/{3}\;) & \cos ({2\pi }/{3}\;) & 0  \\
   0 & 0 & 1  \\
\end{matrix} \right),
\end{align}
\begin{align}
  \mathcal{C}_{2y}=\left(\begin{matrix}
   -1 & 0 & 0  \\
   0 & 1 & 0  \\
   0 & 0 & -1  \\
\end{matrix} \right),
\end{align}
\begin{align}
\mathcal{M}_{xy}=\left( \begin{matrix}
   1 & 0 & 0  \\
   0 & 1 & 0  \\
   0 & 0 & -1  \\
\end{matrix} \right).
\end{align}
Under these symmetry operators, $\mathbf{D}$ transforms as according to Eq. (\ref{Dsymm}):
\begin{widetext}
\begin{align}
\det (\mathcal{C}_{3z}){\mathcal{C}_{3z}}\mathbf{D}\mathcal{C}_{3z}^{-1}=\left( \begin{matrix}
   \frac{1}{4}\left( {{D}_{xx}}+3{{D}_{yy}} \right)+\frac{\sqrt{3}}{4}\left( {{D}_{yx}}+{{D}_{xy}} \right) & \frac{\sqrt{3}}{4}\left( -{{D}_{xx}}+{{D}_{yy}} \right)+\frac{1}{4}\left( {{D}_{xy}}-3{{D}_{yx}} \right) & -\frac{{{D}_{xz}}}{2}-\frac{\sqrt{3}{{D}_{yz}}}{2}  \\
   \frac{\sqrt{3}}{4}\left( -{{D}_{xx}}+{{D}_{yy}} \right)+\frac{1}{4}\left( -3{{D}_{xy}}+{{D}_{yx}} \right) & \frac{1}{4}\left( 3{{D}_{xx}}+{{D}_{yy}} \right)-\frac{\sqrt{3}}{4}\left( {{D}_{yx}}+{{D}_{xy}} \right) & \frac{\sqrt{3}{{D}_{xz}}}{2}-\frac{{{D}_{yz}}}{2}  \\
   -\frac{{{D}_{zx}}}{2}-\frac{\sqrt{3}{{D}_{zy}}}{2} & \frac{\sqrt{3}{{D}_{zx}}}{2}-\frac{{{D}_{zy}}}{2} & {{D}_{zz}}  \\
\end{matrix} \right),	
\label{EqC3}
\end{align}
\end{widetext}
\begin{align}
\det ({\mathcal{C}_{2y}}){\mathcal{C}_{2y}}\mathbf{D}\mathcal{C}_{2y}^{-1}=\left( \begin{matrix}
{{D}_{xx}} & -{{D}_{xy}} & {{D}_{xz}} \\
-{{D}_{yx}} & {{D}_{yy}} & -{{D}_{yz}} \\
{{D}_{zx}} & -{{D}_{zy}} & {{D}_{zz}} \\
\end{matrix} \right),
\label{EqC2y}
\end{align}
\begin{align}
\det ({\mathcal{M}_{xy}}){\mathcal{M}_{xy}}\mathbf{D}\mathcal{M}_{xy}^{-1}=\left( \begin{matrix}
-{{D}_{xx}} & -{{D}_{xy}} & {{D}_{xz}} \\
-{{D}_{yx}} & -{{D}_{yy}} & {{D}_{yz}} \\
{{D}_{zx}} & {{D}_{zy}} & -{{D}_{zz}} \\
\end{matrix} \right).
\label{EqMxy}
\end{align}
The $\mathcal{C}_{3z}$ symmetry ensures the vanish of ${{D}_{xz}}$ and ${{D}_{yz}}$ according to Eqs. (\ref{Dsymm}) and (\ref{EqC3}). Besides, the others $\mathbf{D}$ elements are all zeros under $D_{3h}$ symmetry by combination Eqs. (\ref{Dsymm}) and (\ref{EqC3})-(\ref{EqMxy}).
In other words, there is no $\mathbf{D}$ in 1$H$-MX$_2$ (M=Nb, Ta; X=S, Se) without strain.

When armchair or zigzag strain is applied, the ${\mathcal{C}_{3z}}$ rotation symmetry is broken; however, the ${\mathcal{M}_{xy}}$, ${\mathcal{C}_{2y}}$ and ${\mathcal{M}_{yz}}$ are preserved [Fig. \ref{bandstru} (a)], leading to the ${C}_{2v}$ point group symmetry. That is to say, Eqs. (\ref{EqC2y}) and (\ref{EqMxy}) still satisfy the Eq. (\ref{Dsymm}), while Eq. (\ref{EqC3}) do not. We can get the symmetry of $\mathbf{D}$ under ${{C}_{2v}}$:
\begin{align}
\mathbf{D}=\left( \begin{matrix}
   0 & 0 & {{D}_{xz}}  \\
   0 & 0 & 0  \\
   {{D}_{zx}} & 0 & 0  \\
\end{matrix} \right).
\label{D_C2v}
\end{align}	
${\mathcal{C}_{2y}}$ and ${\mathcal{M}_{yz}}$ guarantee the vanishing of ${{D}_{yz}}$, however ${{D}_{xz}}$ is non-zero under ${{C}_{2v}}$ symmetry. Therefore, the switching of the finite and zero $\mathbf{D}$ can be realized in a metallic 1$H$-MX$_2$ by applying and releasing uniaxial strain. With the appearance of the finite Berry curvature dipole, the nonlinear Hall effect can exist in these 2D non-magnetic metals. These materials may be used as new piezotronics devices.
\begin{table}[!htbp]
\centering
\caption{The comparison electric dipole $\mathbf{P}$ with Berry curvature dipole $\mathbf{D}$.}
\label{Tab1}
\begin{tabular}{p{2.5cm}|p{2.8cm}|p{2.8cm}}
   \hline
   \hline
  & $\mathbf{P}$  & $\mathbf{D}$\\
   \hline
Tensor type &  polar vector  & axial tensor \\
   \hline
Existing space & real space & reciprocal space  \\
   \hline
Existing material  &  insulators  &  metals     \\
   \hline
Direction & parallel to  polar axis & perpendicular to polar axis \\
   \hline
Symmetery constraint &  Eq. (\ref{Psymm}) & Eq. (\ref{Dsymm}) \\
   \hline
Resulting phenomenon & surface charge accumulation

(piezoelectricity) &  nonlinear Hall effect (piezoelectricity) \\
\hline
\hline
\end{tabular}
\end{table}

The method to fabricate the piezotronics devices utilizing $\mathbf{D}$ is different from the conventional one based on $\mathbf{P}$ \cite{RN2232, RN2337, RN2239,RN2231, RN2231, RN2240, RN2345, RN2342}. For example, from above symmetry constrain $\mathbf{D}$ (${{D}_{xz}}$) is perpendicular to the polar axis (${\mathcal{C}_{2y}}$), opposite to $\mathbf{P}$ which is along the polar axis. The difference between $\mathbf{D}$ and $\mathbf{P}$ is because that $\mathbf{P}$ is a polar vector, while $\mathbf{D}$ is an axial tensor (or axial vector in 2D). In contrast to Eq. (\ref{Dsymm}), $\mathbf{P}$ is determined by:
\begin{align}
	\mathbf{P}=S\mathbf{P}.
\label{Psymm}
\end{align}
The differences between Berry curvature dipole and electric dipole are summarized in Table \ref{Tab1}.

\section{Calculation details}
From the above descriptions, the strain can switch $\mathbf{D}$ from zero to finite value in 1$H$-MX$_2$ (M=Nb, Ta; X=S, Se). Thus these materials can in principle be used as piezoelectric devices. Nevertheless, the value of $\mathbf{D}$ and whether it can induce an observable nonlinear Hall effect need to be calculated.

The first-principles calculations based on density functional theory (DFT) were performed by using the QUANTUM-ESPRESSO package \cite{RN82}. Ultrasoft pseudopotentials and general gradient approximation (GGA) according to the Perdew-Burke-Ernzerhof (PBE) functional were used. The energy cutoff of the plane wave (charge density) basis was set to 50 Ry (500 Ry). The Brillouin zone was sampled with a 12$\times$12$\times$1 mesh of $\mathbf{k}$-points. To simulate the monolayer, a 20 \text{\AA} vacuum layer was introduced. The DFT Bloch wave functions were projected to maximally localized Wannier functions by the Wannier90 code \cite{RN149, RN772}. The Berry curvature $\Omega$ and Berry curvature dipole $\mathbf{D}$ were calculated by the WannierTools software package \cite{RN1186}. In the Berry curvature dipole calculations, the convergence test was taken, and a $\mathbf{k}$ mesh grids of 600$\times$600 were adopted. The lattice parameters and atomic positions are fully relaxed until the force on each atom is less than $10^{-4}$ eV/\text{\AA}. For the samples with uniaxial strain, the atomic positions are relaxed with the lattice parameter fixed to strain. 

\section{Results and explanations}

\begin{figure}[h]
\begin{flushleft}
\centering
\includegraphics[width=1\columnwidth]{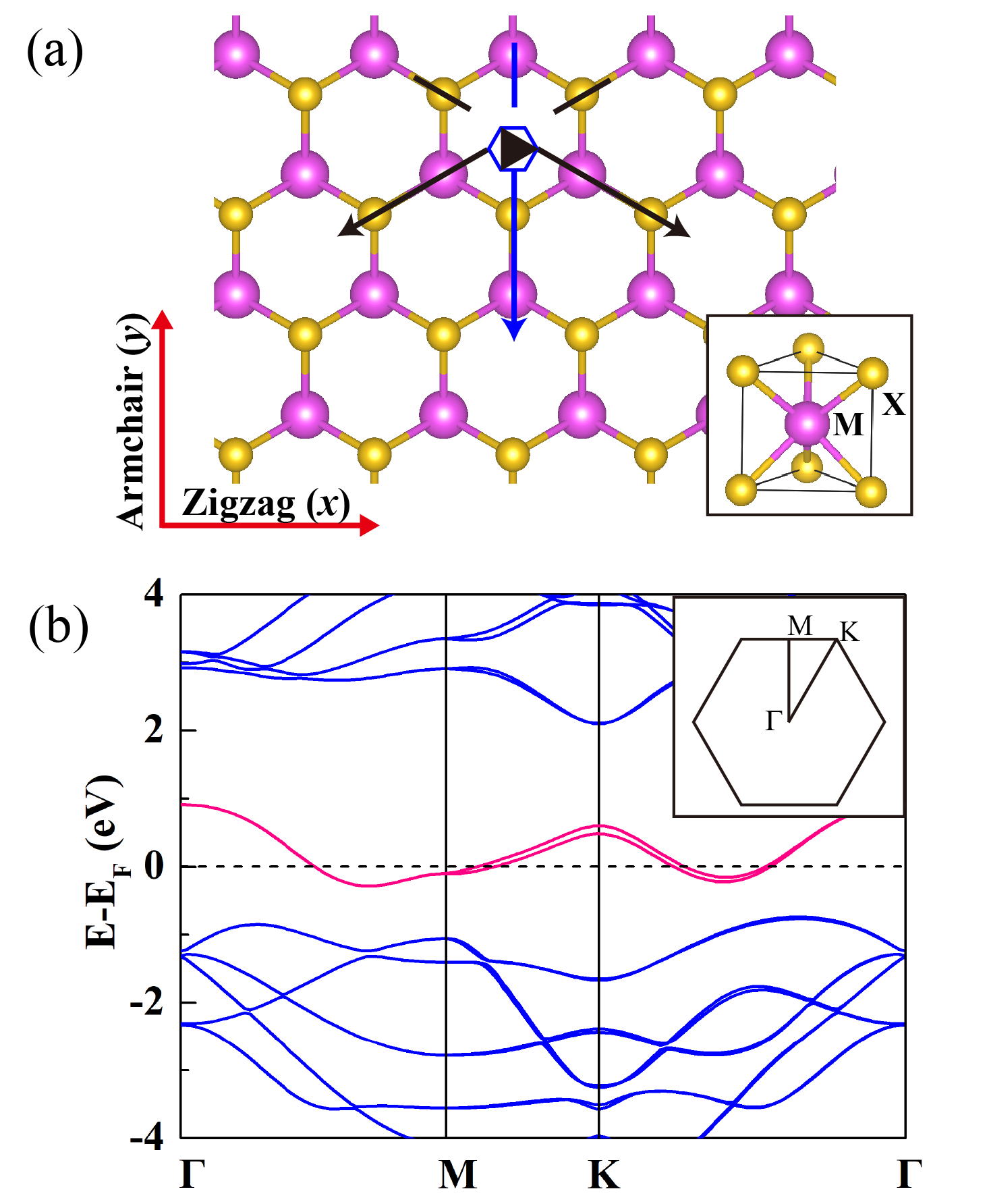}
\caption{(a) Crystal structure of monolayer 1$H$-MX$_2$ (M=Nb, Ta; X=S, Se). The arrows represent the in-plane ${\mathcal{C}_{2}}$ symmetries, line segments represent vertical mirror symmetries and hollowed hexagon denotes for the ${\mathcal{C}_{3z}}$ and ${\mathcal{M}_{xy}}$ symmetries. The symmetries in black (blue) are broken (unbroken) when uniaxial strain is applied. (b) Band structure of monolayer 1$H$-NbS$_2$. The inset illustrates the Brillouin zone.}
\label{bandstru}
\end{flushleft}
\end{figure}

\begin{figure*}
\begin{flushleft}
\centering
\includegraphics[width=1\textwidth]{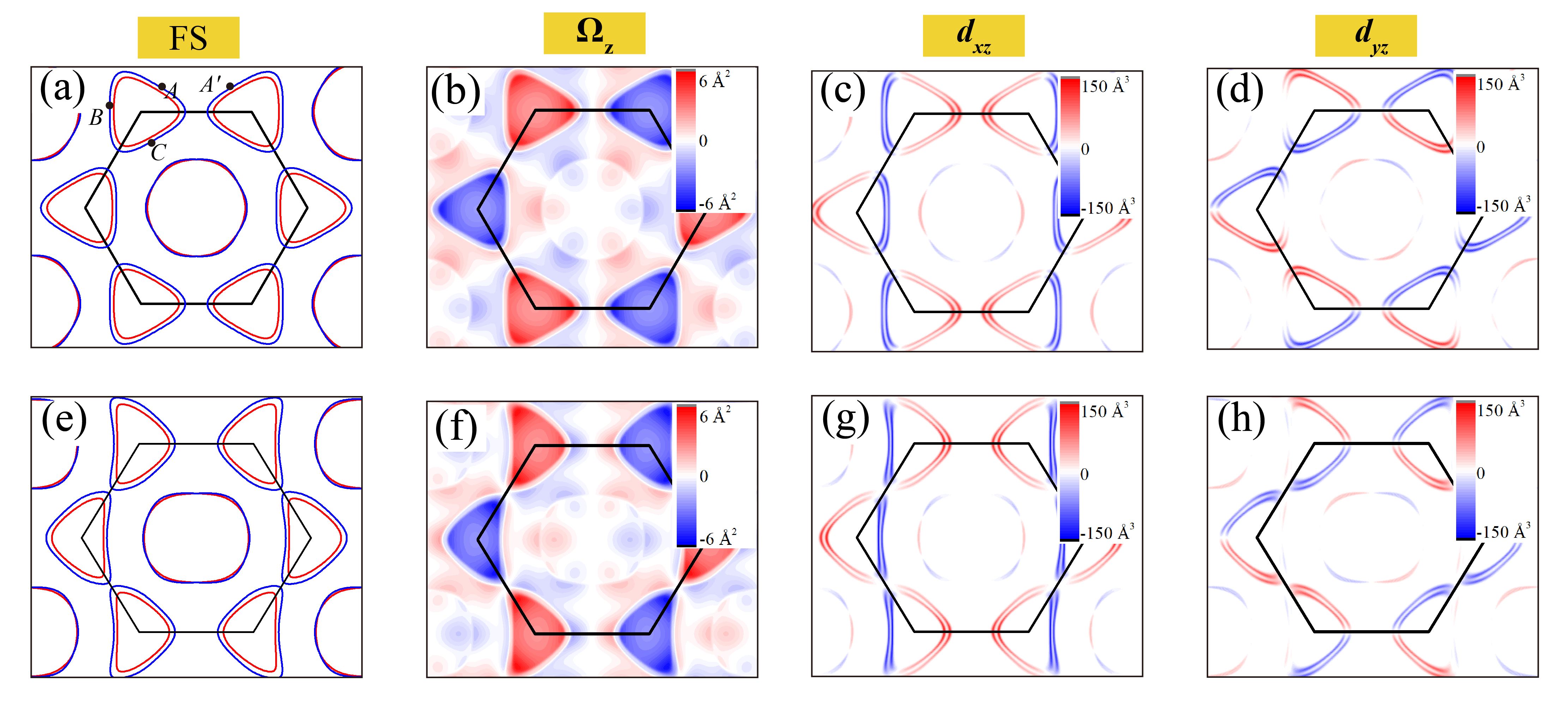}
\caption{(a-d) Upper panel: (a) Fermi surfaces, (b) Berry curvature ${{\Omega }_{z}}(\mathbf{k})$, (c) Berry curvature density ${{d}_{xz}}(\mathbf{k})$  and (d) ${{d}_{yz}}(\mathbf{k})$ of pristine 1$H$-NbS$_2$.  Lower panel: (e-h) are same as (a-d) with a tensile strain of $\xi$=4\% along the armchair direction.}
\label{dipole}
\end{flushleft}
\end{figure*}

Monolayer 1$H$-MX$_2$ (M=Nb, Ta; X=S, Se) were successfully synthesized in experiment recently \cite{RN2204, RN1489, RN1470, RN1540,RN1521, RN1522}. In these materials, one M atom and the nearest six X atoms compose a trigonal prism as shown in Fig. \ref{bandstru} (a), forming a structure with $P\bar{6}m2$ space group (${{D}_{3h}}$ point group symmetry) without inversion symmetry. There is one less $d$ electron in the M atom compared with 1$H$-MoX$_2$ and WX$_2$, leading the metallic ground states. Nonzero Berry curvature dipole density $d_{bd}(\mathbf{k})$ can be introduced without additional doping due to the existence of finite Fermi surface, unlike the semiconductors 1$H$-MoS$_2$ and 1$H$-WSe$_2$ \cite{RN2111, RN2113, RN2172}. These four materials show similar band structures, and here we use 1$H$-NbS$_2$ as a representative.

\begin{figure*}
\begin{flushleft}
\centering
\includegraphics[width=0.8\textwidth]{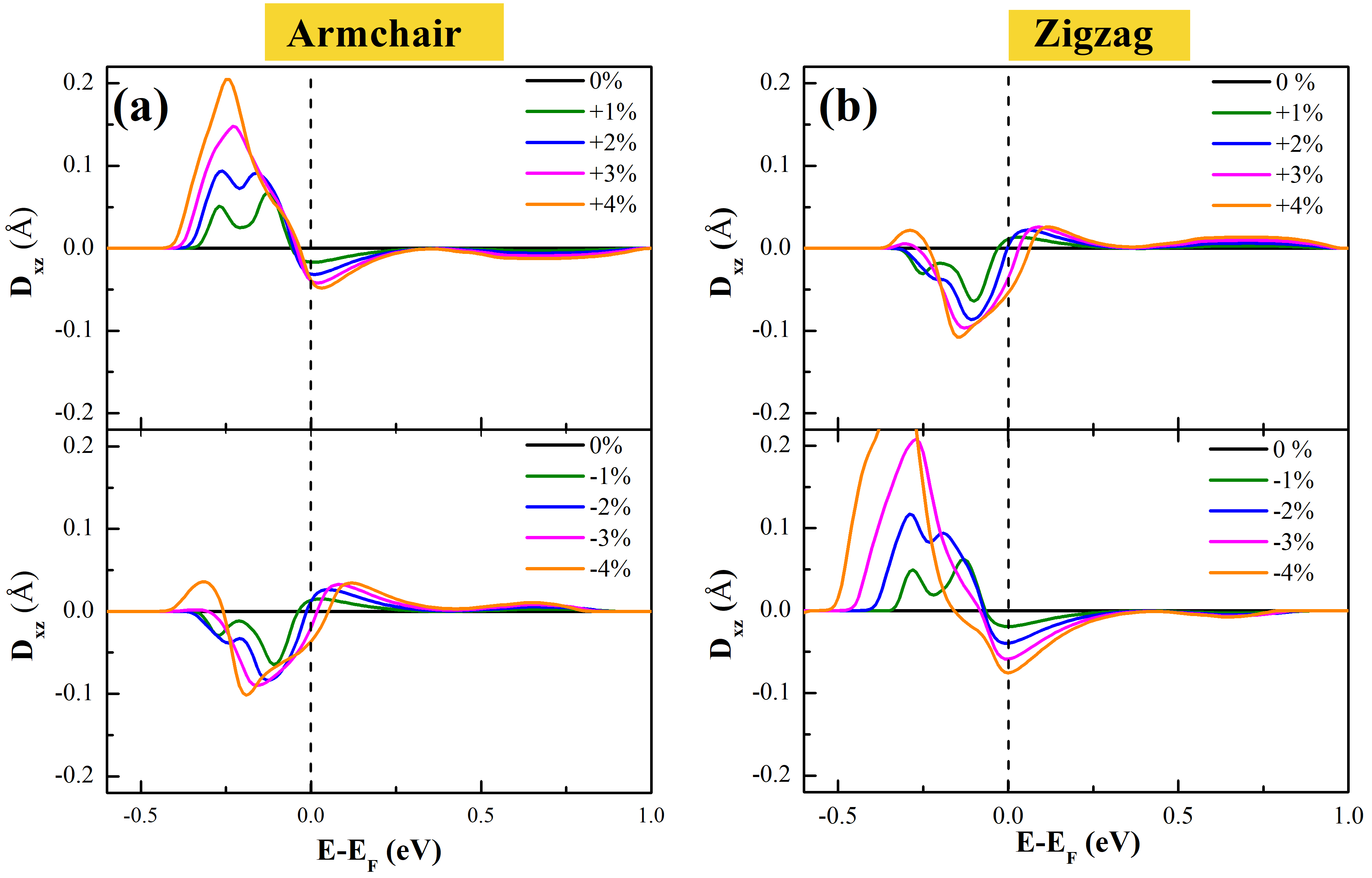}
\caption{${{D}_{xz}}$ of monolayer 1$H$-NbS$_2$ under (a) armchair (b) zigzag strain, for which the positive (negative) strain value represents the tensile (compressive) strain}
\label{Fig-strain}
\end{flushleft}
\end{figure*}

Figure \ref{bandstru}(b) shows the band structure of pristine 1$H$-NbS$_2$. With considering of spin-orbit coupling, there are two half-filled bands cross the Fermi level (${{E}_{F}}$). The largest spin-orbit gap of 116 meV between these two bands is found at the $K$ point. These bands form the Fermi surfaces as shown in Fig. \ref{dipole}(a), where two hexagonal hole pockets located at the $\Gamma$ point and two rounded triangular hole pockets at the $K$ point. The split of two hole pockets around the $K$ point is much larger than that around the $\Gamma$ point. Figures \ref{dipole}(b) exhibits the distribution of Berry curvature ${{\Omega }_{z}}$ of the occupied bands in the Brillouin zone. As discussed in Section II, ${{\Omega }_{z}}$ is odd under the time-reversal symmetry, \emph{i.e.} ${{\Omega }_{z}}(\mathbf{k})=-{{\Omega }_{z}}(-\mathbf{k})$. The largest ${{\Omega }_{z}}$ locates around the $K$ point.

Figures \ref{dipole}(c,d) plots the Berry curvature dipole density ${{d}_{xz}}$ and ${{d}_{yz}}$. Since the Berry curvature dipole is a Fermi surface dependent quantity, ${{d}_{xz}}$ and ${{d}_{yz}}$ are nonzero only on the Fermi lines [see Figs. \ref{dipole}(c), (d)]. We find the largest values of ${{d}_{xz}}$ and ${{d}_{yz}}$ are around the Fermi pockets surrounding $K$ point as well. $d_{xz}$ and $d_{yz}$ is even under time-reversal symmetry [Eq. (\ref{Eq_dbd})], as shown in Figs. \ref{dipole}(c,d). However, these nonzero ${{d}_{xz}}$ and ${{d}_{yz}}$ are not sufficient to lead the emerging of $\mathbf{D}$, due to the presence of the ${\mathcal{C}_{3z}}$ symmetry as disscussed in Section II. It can also be simply explained by comparing ${{\Omega}_{z}}$ and velocities ${{v}_{x}}$ and ${{v}_{y}}$ under the ${\mathcal{C}_{3z}}$ symmetry. The $\mathcal{C}_{3z}$ rotation symmetry can transform the point $A$ in the Fermi surface to points $B$ and $C$, as shown in Fig. \ref{bandstru} (a). $\mathcal{C}_{3z}$ doesn't change the value of  ${{\Omega }_{z}}$ of these three points, \emph{i.e.}, ${{\Omega }_{z}}(A)={{\Omega }_{z}}(B)={{\Omega }_{z}}(C)$. However, $\mathcal{C}_{3z}$ alter the velocities, leading to : $\mathbf{v}(A)+\mathbf{v}(B)+\mathbf{v}(C)=0$ (see Fig. \ref{dbdC3} in the Appendix A). Therefore the integration of the $d_{xz}$ and $d_{yz}$ over the Brillouin zone leads to the vanishing ${{D}_{xz}}$ and ${{D}_{yz}}$ according to Eq. (\ref{EqDbd}). The zero $\mathbf{D}$ of the freestanding materials is important to piezotronics devices because at this case symmetry breaking can switch $\mathbf{D}$ obviously. 

Uniaxial strain (along armchair or zigzag direction) can reduce the symmetry of 1$H$-MX$_2$ to point group ${{C}_{2v}}$. We show the Fermi surfaces, corresponding Berry curvature $\Omega_{z}$, ${{d}_{xz}}(\mathbf{k})$ and ${{d}_{yz}}(\mathbf{k})$ under the tensile strain of $\xi$=4\% along the armchair direction in Figs. \ref{dipole}(e-h), respectively. We find the shapes of Fermi pockets are significantly changed [Fig. \ref{dipole}(e)]. When the $\mathcal{C}_{3z}$ rotation symmetry is broken, $\mathbf{v}\left( A \right)+\mathbf{v}\left( B \right)+\mathbf{v}\left( C \right)\ne 0$. However, the ${{D}_{yz}}$ is still vanishing. This is due to the preserved $\mathcal{C}_{2y}$ symmetry in this point group. $\mathcal{C}_{2y}$ transforms point $A$ to $A'$, and changes the sign of ${{\Omega }_{z}}$, \emph{i.e.} ${{\Omega }_{z}}\left( A \right)=-{{\Omega }_{z}}\left( A' \right)$. Since ${\mathcal{C}_{2y}}$ doesn't influence $v_y$, and ${{v}_{y}}\left( A \right)={{v}_{y}}\left( A' \right)$, ${{D}_{yz}}$ is enforced to be zero, according to Eq. (\ref{EqDbd}). Fortunately ${{D}_{xz}}$ is nonzero under uniaxial strain. The symmetry of of $\mathbf{D}$ from the first-principles calculations agrees with that from macroscopic perspective as Eq. (\ref{D_C2v}). The nonzero $D_{xz}$ means Berry curvature dipole $\mathbf{P}$ along the $x$ direction, \emph{i.e.} zigzag direction, which is different with the electric dipole $\mathbf{P}$ of semiconductors 1$H$-MX$_2$ in piezoelectricity experiments \cite{RN2231, RN2239}. The strain-induced Berry curvature dipole becomes the source of magnetization when there is an in-plane electric field \cite{RN2243}, and it can generate the nonlinear Hall effect in non-magnetic materials.

Figure \ref{Fig-strain} exhibits ${{D}_{xz}}$ of 1$H$-NbS$_2$ under the both armchair and zigzag strain. In the presence of strain, nonzero $D_{xz}$ emerges suddenly and changes linearly with the strain. ${{D}_{xz}}$ is relatively large at the Fermi level, and it can be optimized under small electric gate voltage. As shown in Fig. \ref{Fig-strain}, there are two significant features: (1) ${{D}_{xz}}$ changes its symbol under tensile and compressive strain, and are basically opposite; (2) ${{D}_{xz}}$ under zigzag tensile (compressive) strain shows similar effects with ${{D}_{xz}}$ under the armchair compressive (tensile) strain due to the equivalent impact on the crystal and band structures. Such two features show the dependence of $D_{xz}$ under strain and can be used to manipulate the Berry curvature dipole in the real experiments. We also calculated the Berry curvature dipole under armchair strain for 1$H$-NbSe$_2$, 1$H$-TaS$_2$ and 1$H$-TaSe$_2$ as plotted in Fig. \ref{FigA-Three}. We find the strain effect on ${D}_{xz}$ of these three materials shows similar behavior with that of 1$H$-NbS$_2$.

Next, we estimate the nonlinear Hall signals that $\mathbf{D}$ induced. In 2D materials, the nonlinear Hall current density ${J^{(2\omega )}}$ is expressed as \cite{RN2109} (nonlinear Hall current density ${{J}^{\left( 0 \right)}}$ has a similar form with ${{J}^{\left( 2\omega  \right)}}$)
\begin{align}
\left(\begin{array}{c}{J_{x}^{(2 \omega)}} \\ {J_{y}^{(2 \omega)}}\end{array}\right)=
\left(\begin{array}{cccc}{0} & {\chi_{x x y}} & {0} & {\chi_{x y y}} \\ {\chi_{y x x}} & {0} & {\chi_{y y x}} & {0}\end{array}\right)
\left(\begin{array}{c}{E_{x}^{2}} \\ {E_{x} E_{y}} \\ {E_{y} E_{x}} \\ {E_{y}^{2}}\end{array}\right),
\label{EqOmega}
\end{align}
where $E= \varepsilon {{e}^{i\omega t}}$ is the in-plane driving electric field, $\chi_{a b c}$ is the nonlinear Hall coefficient
\begin{align}
\left\{\begin{array}{l}{\chi_{x y y}=e^{3} \tau /\left[2 \hbar^{2}(1+i \omega \tau)\right] D_{y z}}, \\
{\chi_{y y x}=-e^{3} \tau /\left[2 \hbar^{2}(1+i \omega \tau)\right] D_{y z}}, \\
{\chi_{x x y}=e^{3} \tau /\left[2 \hbar^{2}(1+i \omega \tau)\right] D_{x z}}, \\
{\chi_{y x x}=-e^{3} \tau /\left[2 \hbar^{2}(1+i \omega \tau)\right] D_{x z}},\end{array}\right.
\label{EqOmega}
\end{align}
here $-e$ is the electron charge, and $\tau$ is the relaxation time. In uniaxially strained 1$H$-MX$_2$ (M=Nb, Ta; X=S, Se), only ${{\chi }_{xxy}}$ and ${{\chi }_{yxx}}$ are non-vanishing because of the nonzero ${{D}_{xz}}$. The second-order nonlinear Hall current density and the corresponding nonlinear Hall voltage perpendicular to the driving electric current are
\begin{align}
\left\{ \begin{array}{l}
   J_{\bot }^{\left( 2\omega  \right)}=\frac{{{\mathbf{J}}^{\left( 2\omega  \right)}}\times {{\mathbf{J}}^{\left( \omega  \right)}}}{\left| {{\mathbf{J}}^{\left( \omega  \right)}} \right|}=\frac{{{e}^{3}}\tau }{2{{\hbar }^{2}}\left( 1+i\omega \tau  \right)}{{\left| E \right|}^{2}}{{D}_{xz}}\cos \theta ,  \\
   V_{\bot }^{\left( 2\omega  \right)}=\frac{{{e}^{3}}\tau }{2{{\hbar }^{2}}\left( 1+i\omega \tau  \right)}\rho {{\left| E \right|}^{2}}{{D}_{xz}}l\cos \theta,  \\
\end{array} \right.
\label{NHEsignal}
\end{align}
where $\theta $ means the angle between the driving electric current and zigzag direction (\emph{i.e.} direction of $\mathbf{D}$), $\rho$ means the resistivity, and $l$ refers to the transverse length of the sample. The nonlinear Hall current is proportion to ${\left| E \right|}^{2}{D}_{xz}\cos \theta$ [Eq. (\ref{NHEsignal})]. It means that the nonlinear Hall voltage is maximum when the incident current is along the zigzag direction and vanishes when the incident current is along the armchair direction. The relationships between the strain, Berry curvature dipole, and nonlinear Hall current are summarized in Fig. \ref{BCDpolar}.
\begin{figure}
\begin{flushleft}
\centering
\includegraphics[width=1.0\columnwidth]{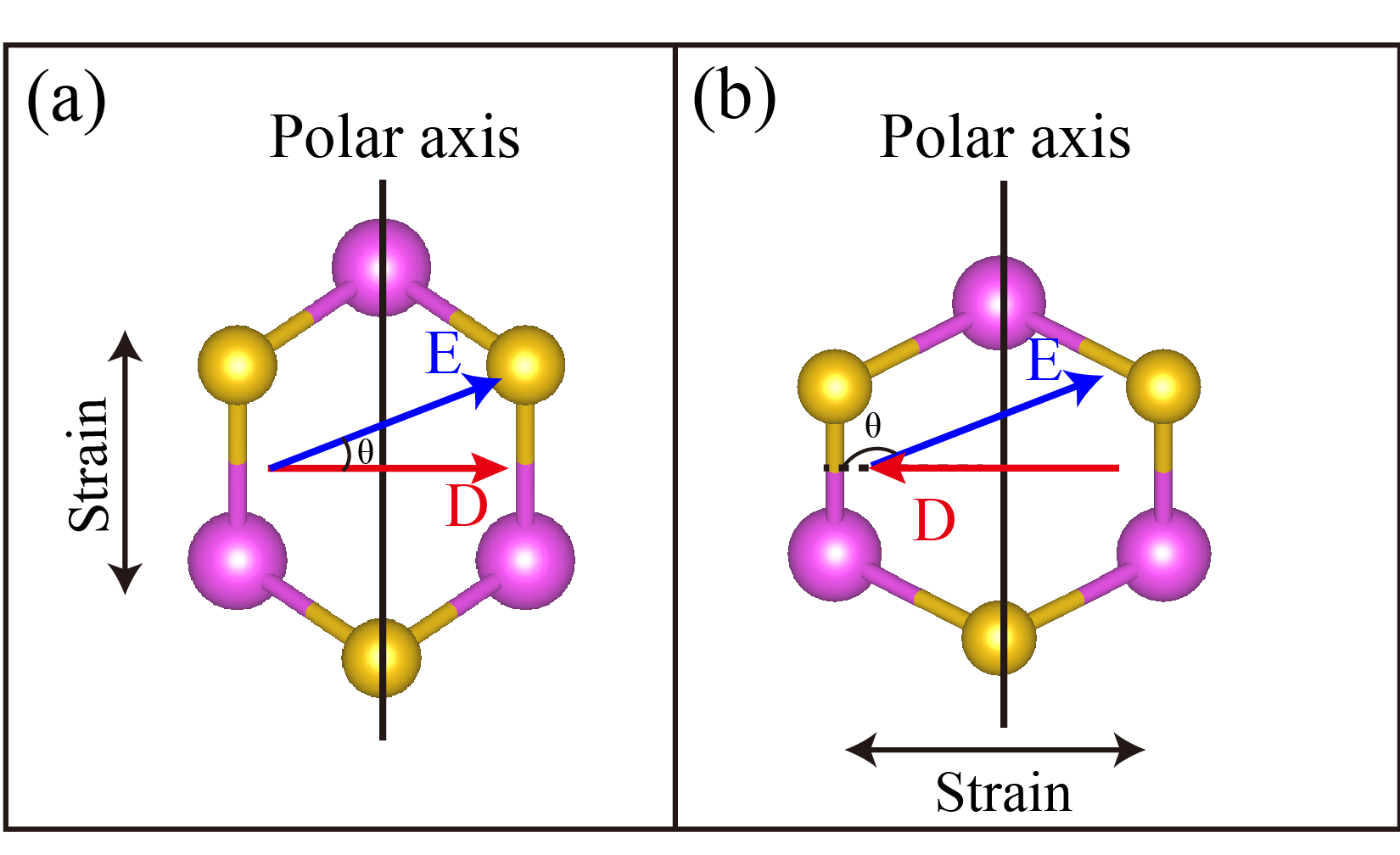}
\caption{The relationships between the strain, Berry curvature dipole $\mathbf{D}$, and nonlinear Hall current. $\mathbf{D}$ is always perpendicular to the polar axis under (a) armchair and (b) zigzag strain, while inverse its direction in two kinds of strain. The nonlinear Hall current is proportional to ${\left| E \right|}^{2}{D}_{xz}\cos \theta$, so the nonlinear Hall effect is maximum when the charge current is parallel to $\mathbf{D}$.}
\label{BCDpolar}
\end{flushleft}
\end{figure}

Taking typical driving current density ${{J}^{\left( \omega  \right)}} \sim$ 10 A/m  \cite{RN2105}, resistivity $\rho \sim {{10}^{4}}$ $\Omega $ normalized from the bulk resistivity in 1$H$-MX$_2$ (M=Nb, Ta; X=S, Se) \cite{RN2253},  relaxation time $\tau \sim 10^{-14}$ s, and transverse length of the sample $l$ = 10 $\mu $m, strain-induced ${{D}_{xz}}\sim 0.05$ $\text{\AA}$ (\emph{e.g.} see Fig. \ref{Fig-strain}) can generate Hall voltage of ~10 $\mu V$, which is comparable to those measured in 2D WTe$_2$ \cite{RN2105, RN2110}. Since strain can switch the nonlinear Hall voltage from zero to obserable value, 1$H$-MX$_2$ (M=Nb, Ta; X=S, Se) can be used in piezotronics devices.

\section{Dissucsion and Conclusion}

The method to fabricate the piezotronics devices utilizing $\mathbf{D}$ is quite different from the traditional one based on $\mathbf{P}$ \cite{RN2232, RN2337, RN2239,RN2231, RN2231, RN2240, RN2345, RN2342}. Piezoelectricity induced by Berry curvature dipole $\mathbf{D}$ is only presented in metals. In comparison, traditional piezoelectricity is presented in insulators.
In the conventional piezotronics devices induced by electric dipole $\mathbf{P}$, the measured signal has the same frequency with the driving field and is sensitive to the disorder variation by strain. Our proposal is based on the nonlinear Hall effect. Similar to nonlinear optics, nonlinear Hall effect doubles the frequency of electrical signals [Eq. (\ref{EqOmega})]. Therefore, it is very easy to separate the response signals from the driving electric current.
Moreover, the piezoelectricity effect origins from Berry curvature dipole, which is highly related to the topological properties of the Fermi surface \cite{RN2111}, thus the piezoelectricity signal is robust against the disorder.
These are advantages in the piezotronics applications induced by $\mathbf{D}$.

Freestanding bilayer 2D WTe$_2$ has a single mirror symmetry ${\mathcal{M}_{yz}}$, and can induce the non-zeros Berry curvature dipole $D_{xz}$ \cite{RN2112, RN2113, RN2105, RN2110, RN2158}. Compared to ${D}_{xz}$ in 1$H$-MX$_2$ (M=Nb, Ta; X=S, Se) induced by uniaxial strain, ${D}_{xz}$ is larger in bilayer WTe$_2$. However, there is no phase transition in bilayer WTe$_2$ when strain is applied, and its ${D}_{xz}$ has no on-off effect. In the view of piezotronics applications, 1$H$-MX$_2$ (M=Nb, Ta; X=S, Se) are more suitable than bilayer 2D WTe$_2$.

In summary, uniaxial strain (along zigzag or armchair direction) can manipulate Berry curvature dipole $\mathbf{D}$ in 2D metals 1$H$-MX$_2$ (M=Nb, Ta; X=S, Se). Without strain, the $\mathcal{C}_{3z}$ symmetry eliminates the $\mathbf{D}$. Uniaxial strain breaks the $\mathcal{C}_{3z}$ symmetry, and guarantees the existing the nonzero $\mathbf{D}$. The direction of $\mathbf{D}$ is perpendicular to the polar axis, different from the relationship between electric dipole $\mathbf{P}$ and polar axis. Uniaxial strain-induced Berry curvature dipole $\mathbf{D}$ can generate observable nonlinear Hall signals under the experimental condition. The maximum nonlinear Hall effect can be obtained when the charge current is parallel to $\mathbf{D}$. The nonlinear Hall signals have different frequencies with the driving current; which has advantages in the piezotronics applications due to the convenience of to be detected. Therefore, 1$H$-MX$_2$ (M=Nb, Ta; X=S, Se) are very promising material platforms for piezotronics devices based on Berry curvature dipole $\mathbf{D}$.
The piezoelectricity utilizing $\mathbf{D}$ in metals is quite different from the traditional one in insulators based on $\mathbf{P}$, due to the nonlinear and topological properties of $\mathbf{D}$. 
We hope our prediction can expend the potential piezotronics application in 2D metallic materials.

\begin{acknowledgments}
This work was supported by the National Nature Science Foundation of China under Grant No. 11822407, the postdoctoral science foundation No. 2018M640513, and a Project Funded by the Priority Academic Program Development of Jiangsu Higher Education Institutions.

\end{acknowledgments}

\appendix
\section{$d_{xz}$ and $d_{yz}$ under $\mathcal{C}_{3z}$  symmetry}
\renewcommand\thefigure{A\arabic{figure}}
\begin{figure}[h]
\begin{flushleft}
\centering
\includegraphics[width=1\columnwidth]{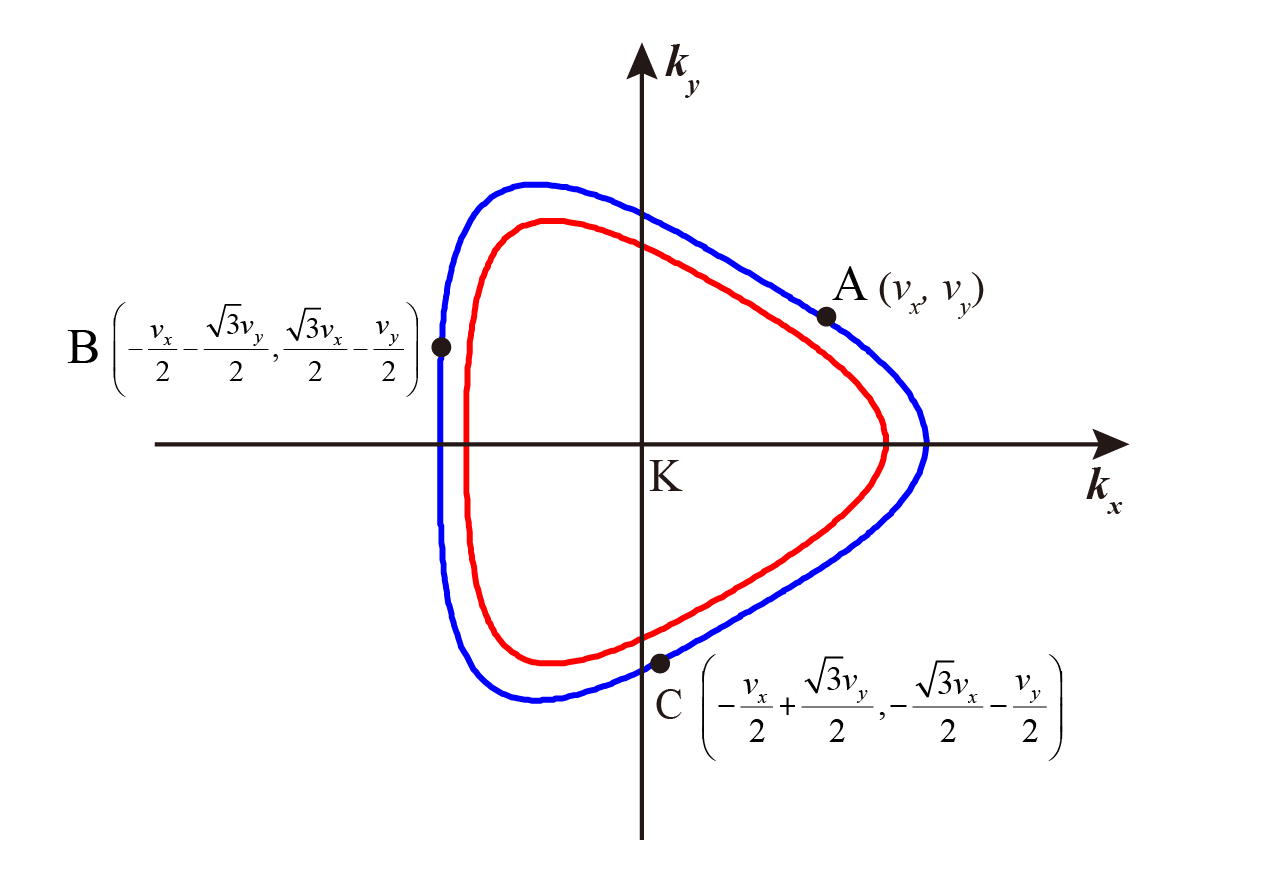}
\setcounter{figure}{0}
\caption{The velocity relationships between the symmetry points around the $K$ point under the $\mathcal{C}_{3z}$ symmetry.}
\label{dbdC3}
\end{flushleft}
\end{figure}

The $\mathcal{C}_{3z}$ rotation symmetry transforms the point $A$ in the Fermi surface to points $B$ and $C$, as shown in Fig. \ref{dbdC3}. ${{\Omega }_{z}}(A)={{\Omega }_{z}}(B)={{\Omega }_{z}}(C)$ and $\mathbf{v}(A)+\mathbf{v}(B)+\mathbf{v}(C)=0$ due to the $\mathcal{C}_{3z}$ symmetry. Therefore the integration of the $d_{xz}$ and $d_{yz}$ over the Brillouin zone leads to the vanishing ${{D}_{xz}}$ and ${{D}_{yz}}$ due to $[v_x(A)\Omega_{z}(A)+v_x(B)\Omega_{z}(B)+v_x(C)\Omega_{z}(C)]\Omega_z=0$ and $[v_y(A)\Omega_{z}(A)+v_y(B)\Omega_{z}(B)+v_y(C)\Omega_{z}(C)]\Omega_z=0$.

\section{${{D}_{xz}}$ of 1$H$-$\text{NbSe}_2$, 1$H$-$\text{TaS}_2$ and 1$H$-$\text{TaSe}_2$ under armchair strain}
\renewcommand\thefigure{B\arabic{figure}}
The Berry curvature dipole under armchair strain for 1$H$-NbSe$_2$, 1$H$-TaS$_2$ and 1$H$-TaSe$_2$ are plotted in  Fig. \ref{FigA-Three}.  $D_{xz}$ changes linearly with strain, and the strain effects on ${D}_{xz}$ of 1$H$-NbSe$_2$ and the 1$H$-TaSe$_2$ are more significant than that of 1$H$-NbS$_2$ and 1$H$-TaS$_2$. The Berry curvature relates to the competition between the spin-orbit coupling effect and bandgap. The 1$H$-NbSe$_2$ can reach a more significant strain effect due to the smaller gap between lower valence bands, even though the spin-orbit coupling is not the strongest.

\begin{figure}[htb]
\centering
\includegraphics[width=1\columnwidth]{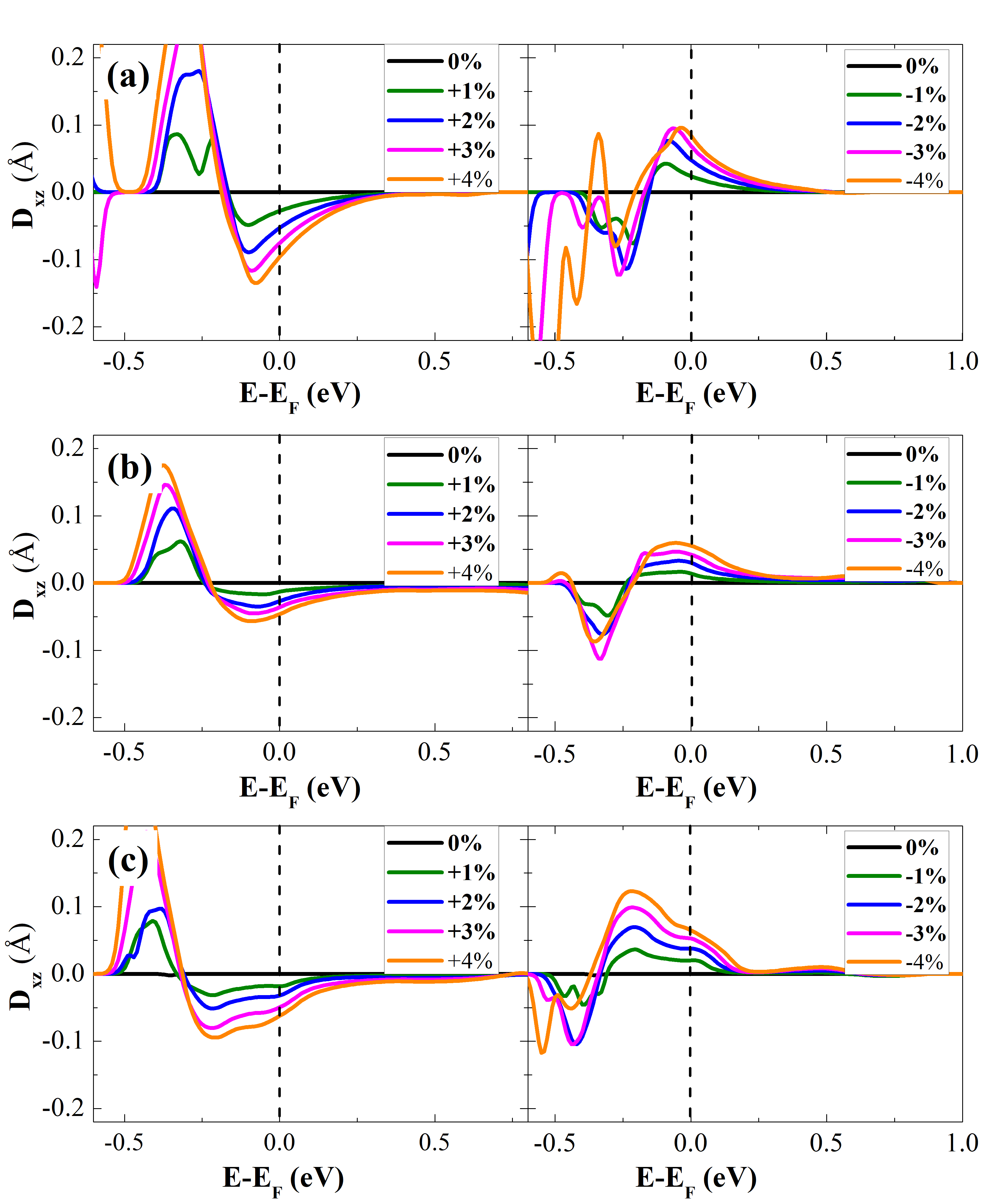}
\setcounter{figure}{0}
\caption{${{D}_{xz}}$ of (a) 1$H$-NbSe$_2$, (b) 1$H$-TaS$_2$ and (c) 1$H$-TaSe$_2$ under different armchair strains.}
\label{FigA-Three}
\end{figure}


\begin{thebibliography}{39}%
\makeatletter
\providecommand \@ifxundefined [1]{%
 \@ifx{#1\undefined}
}%
\providecommand \@ifnum [1]{%
 \ifnum #1\expandafter \@firstoftwo
 \else \expandafter \@secondoftwo
 \fi
}%
\providecommand \@ifx [1]{%
 \ifx #1\expandafter \@firstoftwo
 \else \expandafter \@secondoftwo
 \fi
}%
\providecommand \natexlab [1]{#1}%
\providecommand \enquote  [1]{``#1''}%
\providecommand \bibnamefont  [1]{#1}%
\providecommand \bibfnamefont [1]{#1}%
\providecommand \citenamefont [1]{#1}%
\providecommand \href@noop [0]{\@secondoftwo}%
\providecommand \href [0]{\begingroup \@sanitize@url \@href}%
\providecommand \@href[1]{\@@startlink{#1}\@@href}%
\providecommand \@@href[1]{\endgroup#1\@@endlink}%
\providecommand \@sanitize@url [0]{\catcode `\\12\catcode `\$12\catcode
  `\&12\catcode `\#12\catcode `\^12\catcode `\_12\catcode `\%12\relax}%
\providecommand \@@startlink[1]{}%
\providecommand \@@endlink[0]{}%
\providecommand \url  [0]{\begingroup\@sanitize@url \@url }%
\providecommand \@url [1]{\endgroup\@href {#1}{\urlprefix }}%
\providecommand \urlprefix  [0]{URL }%
\providecommand \Eprint [0]{\href }%
\providecommand \doibase [0]{http://dx.doi.org/}%
\providecommand \selectlanguage [0]{\@gobble}%
\providecommand \bibinfo  [0]{\@secondoftwo}%
\providecommand \bibfield  [0]{\@secondoftwo}%
\providecommand \translation [1]{[#1]}%
\providecommand \BibitemOpen [0]{}%
\providecommand \bibitemStop [0]{}%
\providecommand \bibitemNoStop [0]{.\EOS\space}%
\providecommand \EOS [0]{\spacefactor3000\relax}%
\providecommand \BibitemShut  [1]{\csname bibitem#1\endcsname}%
\let\auto@bib@innerbib\@empty
\bibitem [{\citenamefont {Fei}\ \emph {et~al.}(2009)\citenamefont {Fei},
  \citenamefont {Yeh}, \citenamefont {Zhou}, \citenamefont {Xu}, \citenamefont
  {Gao}, \citenamefont {Song}, \citenamefont {Gu}, \citenamefont {Huang},\ and\
  \citenamefont {Wang}}]{RN2248}%
  \BibitemOpen
  \bibfield  {author} {\bibinfo {author} {\bibfnamefont {P.}~\bibnamefont
  {Fei}}, \bibinfo {author} {\bibfnamefont {P.~H.}\ \bibnamefont {Yeh}},
  \bibinfo {author} {\bibfnamefont {J.}~\bibnamefont {Zhou}}, \bibinfo {author}
  {\bibfnamefont {S.}~\bibnamefont {Xu}}, \bibinfo {author} {\bibfnamefont
  {Y.}~\bibnamefont {Gao}}, \bibinfo {author} {\bibfnamefont {J.}~\bibnamefont
  {Song}}, \bibinfo {author} {\bibfnamefont {Y.}~\bibnamefont {Gu}}, \bibinfo
  {author} {\bibfnamefont {Y.}~\bibnamefont {Huang}}, \ and\ \bibinfo {author}
  {\bibfnamefont {Z.~L.}\ \bibnamefont {Wang}},\ }\href {\doibase
  10.1021/nl901606b} {\bibfield  {journal} {\bibinfo  {journal} {Nano Lett.}\
  }\textbf {\bibinfo {volume} {9}},\ \bibinfo {pages} {3435} (\bibinfo {year}
  {2009})}\BibitemShut {NoStop}%
\bibitem [{\citenamefont {He}\ \emph {et~al.}(2007)\citenamefont {He},
  \citenamefont {Hsin}, \citenamefont {Liu}, \citenamefont {Chen},\ and\
  \citenamefont {Wang}}]{RN2245}%
  \BibitemOpen
  \bibfield  {author} {\bibinfo {author} {\bibfnamefont {J.}~\bibnamefont
  {He}}, \bibinfo {author} {\bibfnamefont {C.}~\bibnamefont {Hsin}}, \bibinfo
  {author} {\bibfnamefont {J.}~\bibnamefont {Liu}}, \bibinfo {author}
  {\bibfnamefont {L.}~\bibnamefont {Chen}}, \ and\ \bibinfo {author}
  {\bibfnamefont {Z.}~\bibnamefont {Wang}},\ }\href {\doibase
  10.1002/adma.200601908} {\bibfield  {journal} {\bibinfo  {journal} {Adv.
  Mater.}\ }\textbf {\bibinfo {volume} {19}},\ \bibinfo {pages} {781} (\bibinfo
  {year} {2007})}\BibitemShut {NoStop}%
\bibitem [{\citenamefont {Kingon}\ and\ \citenamefont
  {Srinivasan}(2005)}]{RN2235}%
  \BibitemOpen
  \bibfield  {author} {\bibinfo {author} {\bibfnamefont {A.~I.}\ \bibnamefont
  {Kingon}}\ and\ \bibinfo {author} {\bibfnamefont {S.}~\bibnamefont
  {Srinivasan}},\ }\href {\doibase 10.1038/nmat1334} {\bibfield  {journal}
  {\bibinfo  {journal} {Nat. Mater.}\ }\textbf {\bibinfo {volume} {4}},\
  \bibinfo {pages} {233} (\bibinfo {year} {2005})}\BibitemShut {NoStop}%
\bibitem [{\citenamefont {Liu}\ \emph {et~al.}(2010)\citenamefont {Liu},
  \citenamefont {Lee}, \citenamefont {Ding}, \citenamefont {Liu},\ and\
  \citenamefont {Wang}}]{RN2250}%
  \BibitemOpen
  \bibfield  {author} {\bibinfo {author} {\bibfnamefont {W.}~\bibnamefont
  {Liu}}, \bibinfo {author} {\bibfnamefont {M.}~\bibnamefont {Lee}}, \bibinfo
  {author} {\bibfnamefont {L.}~\bibnamefont {Ding}}, \bibinfo {author}
  {\bibfnamefont {J.}~\bibnamefont {Liu}}, \ and\ \bibinfo {author}
  {\bibfnamefont {Z.~L.}\ \bibnamefont {Wang}},\ }\href {\doibase
  10.1021/nl1017145} {\bibfield  {journal} {\bibinfo  {journal} {Nano Lett.}\
  }\textbf {\bibinfo {volume} {10}},\ \bibinfo {pages} {3084} (\bibinfo {year}
  {2010})}\BibitemShut {NoStop}%
\bibitem [{\citenamefont {Wang}\ \emph {et~al.}(2006)\citenamefont {Wang},
  \citenamefont {Zhou}, \citenamefont {Song}, \citenamefont {Liu},
  \citenamefont {Xu},\ and\ \citenamefont {Wang}}]{RN2251}%
  \BibitemOpen
  \bibfield  {author} {\bibinfo {author} {\bibfnamefont {X.}~\bibnamefont
  {Wang}}, \bibinfo {author} {\bibfnamefont {J.}~\bibnamefont {Zhou}}, \bibinfo
  {author} {\bibfnamefont {J.}~\bibnamefont {Song}}, \bibinfo {author}
  {\bibfnamefont {J.}~\bibnamefont {Liu}}, \bibinfo {author} {\bibfnamefont
  {N.}~\bibnamefont {Xu}}, \ and\ \bibinfo {author} {\bibfnamefont {Z.~L.}\
  \bibnamefont {Wang}},\ }\href {\doibase 10.1021/nl061802g} {\bibfield
  {journal} {\bibinfo  {journal} {Nano Lett.}\ }\textbf {\bibinfo {volume}
  {6}},\ \bibinfo {pages} {2768} (\bibinfo {year} {2006})}\BibitemShut
  {NoStop}%
\bibitem [{\citenamefont {Wang}(2007)}]{RN2244}%
  \BibitemOpen
  \bibfield  {author} {\bibinfo {author} {\bibfnamefont {Z.}~\bibnamefont
  {Wang}},\ }\href {\doibase 10.1002/adma.200602918} {\bibfield  {journal}
  {\bibinfo  {journal} {Adv. Mater.}\ }\textbf {\bibinfo {volume} {19}},\
  \bibinfo {pages} {889} (\bibinfo {year} {2007})}\BibitemShut {NoStop}%
\bibitem [{\citenamefont {Wang}\ and\ \citenamefont {Song}(2006)}]{RN2233}%
  \BibitemOpen
  \bibfield  {author} {\bibinfo {author} {\bibfnamefont {Z.~L.}\ \bibnamefont
  {Wang}}\ and\ \bibinfo {author} {\bibfnamefont {J.}~\bibnamefont {Song}},\
  }\href {\doibase 10.1126/science.1124005} {\bibfield  {journal} {\bibinfo
  {journal} {Science}\ }\textbf {\bibinfo {volume} {312}},\ \bibinfo {pages}
  {242} (\bibinfo {year} {2006})}\BibitemShut {NoStop}%
\bibitem [{\citenamefont {Wu}\ \emph {et~al.}(2010)\citenamefont {Wu},
  \citenamefont {Wei},\ and\ \citenamefont {Wang}}]{RN2249}%
  \BibitemOpen
  \bibfield  {author} {\bibinfo {author} {\bibfnamefont {W.}~\bibnamefont
  {Wu}}, \bibinfo {author} {\bibfnamefont {Y.}~\bibnamefont {Wei}}, \ and\
  \bibinfo {author} {\bibfnamefont {Z.~L.}\ \bibnamefont {Wang}},\ }\href
  {\doibase 10.1002/adma.201001925} {\bibfield  {journal} {\bibinfo  {journal}
  {Adv. Mater.}\ }\textbf {\bibinfo {volume} {22}},\ \bibinfo {pages} {4711}
  (\bibinfo {year} {2010})}\BibitemShut {NoStop}%
\bibitem [{\citenamefont {Yang}\ \emph {et~al.}(2008)\citenamefont {Yang},
  \citenamefont {Qin}, \citenamefont {Dai},\ and\ \citenamefont
  {Wang}}]{RN2234}%
  \BibitemOpen
  \bibfield  {author} {\bibinfo {author} {\bibfnamefont {R.}~\bibnamefont
  {Yang}}, \bibinfo {author} {\bibfnamefont {Y.}~\bibnamefont {Qin}}, \bibinfo
  {author} {\bibfnamefont {L.}~\bibnamefont {Dai}}, \ and\ \bibinfo {author}
  {\bibfnamefont {Z.~L.}\ \bibnamefont {Wang}},\ }\href {\doibase
  10.1038/nnano.2008.314
  https://www.nature.com/articles/nnano.2008.314#supplementary-information}
  {\bibfield  {journal} {\bibinfo  {journal} {Nat. Nanotechnol.}\ }\textbf
  {\bibinfo {volume} {4}},\ \bibinfo {pages} {34} (\bibinfo {year}
  {2008})}\BibitemShut {NoStop}%
\bibitem [{\citenamefont {Zhou}\ \emph {et~al.}(2008)\citenamefont {Zhou},
  \citenamefont {Gu}, \citenamefont {Fei}, \citenamefont {Mai}, \citenamefont
  {Gao}, \citenamefont {Yang}, \citenamefont {Bao},\ and\ \citenamefont
  {Wang}}]{RN2252}%
  \BibitemOpen
  \bibfield  {author} {\bibinfo {author} {\bibfnamefont {J.}~\bibnamefont
  {Zhou}}, \bibinfo {author} {\bibfnamefont {Y.}~\bibnamefont {Gu}}, \bibinfo
  {author} {\bibfnamefont {P.}~\bibnamefont {Fei}}, \bibinfo {author}
  {\bibfnamefont {W.}~\bibnamefont {Mai}}, \bibinfo {author} {\bibfnamefont
  {Y.}~\bibnamefont {Gao}}, \bibinfo {author} {\bibfnamefont {R.}~\bibnamefont
  {Yang}}, \bibinfo {author} {\bibfnamefont {G.}~\bibnamefont {Bao}}, \ and\
  \bibinfo {author} {\bibfnamefont {Z.~L.}\ \bibnamefont {Wang}},\ }\href
  {\doibase 10.1021/nl802367t} {\bibfield  {journal} {\bibinfo  {journal} {Nano
  Lett.}\ }\textbf {\bibinfo {volume} {8}},\ \bibinfo {pages} {3035} (\bibinfo
  {year} {2008})}\BibitemShut {NoStop}%
\bibitem [{\citenamefont {Duerloo}\ \emph {et~al.}(2012)\citenamefont
  {Duerloo}, \citenamefont {Ong},\ and\ \citenamefont {Reed}}]{RN2232}%
  \BibitemOpen
  \bibfield  {author} {\bibinfo {author} {\bibfnamefont {K.-A.~N.}\
  \bibnamefont {Duerloo}}, \bibinfo {author} {\bibfnamefont {M.~T.}\
  \bibnamefont {Ong}}, \ and\ \bibinfo {author} {\bibfnamefont {E.~J.}\
  \bibnamefont {Reed}},\ }\href {\doibase 10.1021/jz3012436} {\bibfield
  {journal} {\bibinfo  {journal} {J. Phys. Chem. Lett.}\ }\textbf {\bibinfo
  {volume} {3}},\ \bibinfo {pages} {2871} (\bibinfo {year} {2012})}\BibitemShut
  {NoStop}%
\bibitem [{\citenamefont {Wu}\ \emph {et~al.}(2014)\citenamefont {Wu},
  \citenamefont {Wang}, \citenamefont {Li}, \citenamefont {Zhang},
  \citenamefont {Lin}, \citenamefont {Niu}, \citenamefont {Chenet},
  \citenamefont {Zhang}, \citenamefont {Hao}, \citenamefont {Heinz},
  \citenamefont {Hone},\ and\ \citenamefont {Wang}}]{RN2231}%
  \BibitemOpen
  \bibfield  {author} {\bibinfo {author} {\bibfnamefont {W.}~\bibnamefont
  {Wu}}, \bibinfo {author} {\bibfnamefont {L.}~\bibnamefont {Wang}}, \bibinfo
  {author} {\bibfnamefont {Y.}~\bibnamefont {Li}}, \bibinfo {author}
  {\bibfnamefont {F.}~\bibnamefont {Zhang}}, \bibinfo {author} {\bibfnamefont
  {L.}~\bibnamefont {Lin}}, \bibinfo {author} {\bibfnamefont {S.}~\bibnamefont
  {Niu}}, \bibinfo {author} {\bibfnamefont {D.}~\bibnamefont {Chenet}},
  \bibinfo {author} {\bibfnamefont {X.}~\bibnamefont {Zhang}}, \bibinfo
  {author} {\bibfnamefont {Y.}~\bibnamefont {Hao}}, \bibinfo {author}
  {\bibfnamefont {T.~F.}\ \bibnamefont {Heinz}}, \bibinfo {author}
  {\bibfnamefont {J.}~\bibnamefont {Hone}}, \ and\ \bibinfo {author}
  {\bibfnamefont {Z.~L.}\ \bibnamefont {Wang}},\ }\href {\doibase
  10.1038/nature13792} {\bibfield  {journal} {\bibinfo  {journal} {Nature}\
  }\textbf {\bibinfo {volume} {514}},\ \bibinfo {pages} {470} (\bibinfo {year}
  {2014})}\BibitemShut {NoStop}%
\bibitem [{\citenamefont {Zhu}\ \emph {et~al.}(2014)\citenamefont {Zhu},
  \citenamefont {Wang}, \citenamefont {Xiao}, \citenamefont {Liu},
  \citenamefont {Xiong}, \citenamefont {Wong}, \citenamefont {Ye},
  \citenamefont {Ye}, \citenamefont {Yin},\ and\ \citenamefont
  {Zhang}}]{RN2342}%
  \BibitemOpen
  \bibfield  {author} {\bibinfo {author} {\bibfnamefont {H.}~\bibnamefont
  {Zhu}}, \bibinfo {author} {\bibfnamefont {Y.}~\bibnamefont {Wang}}, \bibinfo
  {author} {\bibfnamefont {J.}~\bibnamefont {Xiao}}, \bibinfo {author}
  {\bibfnamefont {M.}~\bibnamefont {Liu}}, \bibinfo {author} {\bibfnamefont
  {S.}~\bibnamefont {Xiong}}, \bibinfo {author} {\bibfnamefont {Z.~J.}\
  \bibnamefont {Wong}}, \bibinfo {author} {\bibfnamefont {Z.}~\bibnamefont
  {Ye}}, \bibinfo {author} {\bibfnamefont {Y.}~\bibnamefont {Ye}}, \bibinfo
  {author} {\bibfnamefont {X.}~\bibnamefont {Yin}}, \ and\ \bibinfo {author}
  {\bibfnamefont {X.}~\bibnamefont {Zhang}},\ }\href {\doibase
  10.1038/nnano.2014.309} {\bibfield  {journal} {\bibinfo  {journal} {Nat.
  Nanotechnol.}\ }\textbf {\bibinfo {volume} {10}},\ \bibinfo {pages} {151}
  (\bibinfo {year} {2014})}\BibitemShut {NoStop}%
\bibitem [{\citenamefont {Birkholz}(1995)}]{RN2337}%
  \BibitemOpen
  \bibfield  {author} {\bibinfo {author} {\bibfnamefont {M.}~\bibnamefont
  {Birkholz}},\ }\href {\doibase 10.1007/bf01313055} {\bibfield  {journal}
  {\bibinfo  {journal} {Zeitschrift f邦r Physik B Condensed Matter}\ }\textbf
  {\bibinfo {volume} {96}},\ \bibinfo {pages} {333} (\bibinfo {year}
  {1995})}\BibitemShut {NoStop}%
\bibitem [{\citenamefont {Bertolazzi}\ \emph {et~al.}(2011)\citenamefont
  {Bertolazzi}, \citenamefont {Brivio},\ and\ \citenamefont {Kis}}]{RN2239}%
  \BibitemOpen
  \bibfield  {author} {\bibinfo {author} {\bibfnamefont {S.}~\bibnamefont
  {Bertolazzi}}, \bibinfo {author} {\bibfnamefont {J.}~\bibnamefont {Brivio}},
  \ and\ \bibinfo {author} {\bibfnamefont {A.}~\bibnamefont {Kis}},\ }\href
  {\doibase 10.1021/nn203879f} {\bibfield  {journal} {\bibinfo  {journal} {ACS
  Nano}\ }\textbf {\bibinfo {volume} {5}},\ \bibinfo {pages} {9703} (\bibinfo
  {year} {2011})}\BibitemShut {NoStop}%
\bibitem [{\citenamefont {Lee}\ \emph {et~al.}(2008)\citenamefont {Lee},
  \citenamefont {Wei}, \citenamefont {Kysar},\ and\ \citenamefont
  {Hone}}]{RN2240}%
  \BibitemOpen
  \bibfield  {author} {\bibinfo {author} {\bibfnamefont {C.}~\bibnamefont
  {Lee}}, \bibinfo {author} {\bibfnamefont {X.}~\bibnamefont {Wei}}, \bibinfo
  {author} {\bibfnamefont {J.~W.}\ \bibnamefont {Kysar}}, \ and\ \bibinfo
  {author} {\bibfnamefont {J.}~\bibnamefont {Hone}},\ }\href {\doibase
  10.1126/science.1157996} {\bibfield  {journal} {\bibinfo  {journal}
  {Science}\ }\textbf {\bibinfo {volume} {321}},\ \bibinfo {pages} {385}
  (\bibinfo {year} {2008})}\BibitemShut {NoStop}%
\bibitem [{\citenamefont {Blonsky}\ \emph {et~al.}(2015)\citenamefont
  {Blonsky}, \citenamefont {Zhuang}, \citenamefont {Singh},\ and\ \citenamefont
  {Hennig}}]{RN2345}%
  \BibitemOpen
  \bibfield  {author} {\bibinfo {author} {\bibfnamefont {M.~N.}\ \bibnamefont
  {Blonsky}}, \bibinfo {author} {\bibfnamefont {H.~L.}\ \bibnamefont {Zhuang}},
  \bibinfo {author} {\bibfnamefont {A.~K.}\ \bibnamefont {Singh}}, \ and\
  \bibinfo {author} {\bibfnamefont {R.~G.}\ \bibnamefont {Hennig}},\ }\href
  {\doibase 10.1021/acsnano.5b03394} {\bibfield  {journal} {\bibinfo  {journal}
  {ACS Nano}\ }\textbf {\bibinfo {volume} {9}},\ \bibinfo {pages} {9885}
  (\bibinfo {year} {2015})}\BibitemShut {NoStop}%
\bibitem [{\citenamefont {Sodemann}\ and\ \citenamefont {Fu}(2015)}]{RN2111}%
  \BibitemOpen
  \bibfield  {author} {\bibinfo {author} {\bibfnamefont {I.}~\bibnamefont
  {Sodemann}}\ and\ \bibinfo {author} {\bibfnamefont {L.}~\bibnamefont {Fu}},\
  }\href {\doibase 10.1103/PhysRevLett.115.216806} {\bibfield  {journal}
  {\bibinfo  {journal} {Phys. Rev. Lett.}\ }\textbf {\bibinfo {volume} {115}},\
  \bibinfo {pages} {216806} (\bibinfo {year} {2015})}\BibitemShut {NoStop}%
\bibitem [{\citenamefont {Kang}\ \emph {et~al.}(2019)\citenamefont {Kang},
  \citenamefont {Li}, \citenamefont {Sohn}, \citenamefont {Shan},\ and\
  \citenamefont {Mak}}]{RN2105}%
  \BibitemOpen
  \bibfield  {author} {\bibinfo {author} {\bibfnamefont {K.}~\bibnamefont
  {Kang}}, \bibinfo {author} {\bibfnamefont {T.}~\bibnamefont {Li}}, \bibinfo
  {author} {\bibfnamefont {E.}~\bibnamefont {Sohn}}, \bibinfo {author}
  {\bibfnamefont {J.}~\bibnamefont {Shan}}, \ and\ \bibinfo {author}
  {\bibfnamefont {K.~F.}\ \bibnamefont {Mak}},\ }\href {\doibase
  https://doi.org/10.1038/s41563-019-0294-7} {\bibfield  {journal} {\bibinfo
  {journal} {Nat. Mater.}\ }\textbf {\bibinfo {volume} {18}},\ \bibinfo {pages}
  {324} (\bibinfo {year} {2019})}\BibitemShut {NoStop}%
\bibitem [{\citenamefont {Ma}\ \emph {et~al.}(2019)\citenamefont {Ma},
  \citenamefont {Xu}, \citenamefont {Shen}, \citenamefont {MacNeill},
  \citenamefont {Fatemi}, \citenamefont {Chang}, \citenamefont {Mier~Valdivia},
  \citenamefont {Wu}, \citenamefont {Du}, \citenamefont {Hsu}, \citenamefont
  {Fang}, \citenamefont {Gibson}, \citenamefont {Watanabe}, \citenamefont
  {Taniguchi}, \citenamefont {Cava}, \citenamefont {Kaxiras}, \citenamefont
  {Lu}, \citenamefont {Lin}, \citenamefont {Fu}, \citenamefont {Gedik},\ and\
  \citenamefont {Jarillo-Herrero}}]{RN2110}%
  \BibitemOpen
  \bibfield  {author} {\bibinfo {author} {\bibfnamefont {Q.}~\bibnamefont
  {Ma}}, \bibinfo {author} {\bibfnamefont {S.~Y.}\ \bibnamefont {Xu}}, \bibinfo
  {author} {\bibfnamefont {H.}~\bibnamefont {Shen}}, \bibinfo {author}
  {\bibfnamefont {D.}~\bibnamefont {MacNeill}}, \bibinfo {author}
  {\bibfnamefont {V.}~\bibnamefont {Fatemi}}, \bibinfo {author} {\bibfnamefont
  {T.~R.}\ \bibnamefont {Chang}}, \bibinfo {author} {\bibfnamefont {A.~M.}\
  \bibnamefont {Mier~Valdivia}}, \bibinfo {author} {\bibfnamefont
  {S.}~\bibnamefont {Wu}}, \bibinfo {author} {\bibfnamefont {Z.}~\bibnamefont
  {Du}}, \bibinfo {author} {\bibfnamefont {C.~H.}\ \bibnamefont {Hsu}},
  \bibinfo {author} {\bibfnamefont {S.}~\bibnamefont {Fang}}, \bibinfo {author}
  {\bibfnamefont {Q.~D.}\ \bibnamefont {Gibson}}, \bibinfo {author}
  {\bibfnamefont {K.}~\bibnamefont {Watanabe}}, \bibinfo {author}
  {\bibfnamefont {T.}~\bibnamefont {Taniguchi}}, \bibinfo {author}
  {\bibfnamefont {R.~J.}\ \bibnamefont {Cava}}, \bibinfo {author}
  {\bibfnamefont {E.}~\bibnamefont {Kaxiras}}, \bibinfo {author} {\bibfnamefont
  {H.~Z.}\ \bibnamefont {Lu}}, \bibinfo {author} {\bibfnamefont
  {H.}~\bibnamefont {Lin}}, \bibinfo {author} {\bibfnamefont {L.}~\bibnamefont
  {Fu}}, \bibinfo {author} {\bibfnamefont {N.}~\bibnamefont {Gedik}}, \ and\
  \bibinfo {author} {\bibfnamefont {P.}~\bibnamefont {Jarillo-Herrero}},\
  }\href {\doibase 10.1038/s41586-018-0807-6} {\bibfield  {journal} {\bibinfo
  {journal} {Nature}\ }\textbf {\bibinfo {volume} {565}},\ \bibinfo {pages}
  {337} (\bibinfo {year} {2019})}\BibitemShut {NoStop}%
\bibitem [{\citenamefont {Xu}\ \emph {et~al.}(2018)\citenamefont {Xu},
  \citenamefont {Ma}, \citenamefont {Shen}, \citenamefont {Fatemi},
  \citenamefont {Wu}, \citenamefont {Chang}, \citenamefont {Chang},
  \citenamefont {Valdivia}, \citenamefont {Chan}, \citenamefont {Gibson},
  \citenamefont {Zhou}, \citenamefont {Liu}, \citenamefont {Watanabe},
  \citenamefont {Taniguchi}, \citenamefont {Lin}, \citenamefont {Cava},
  \citenamefont {Fu}, \citenamefont {Gedik},\ and\ \citenamefont
  {Jarillo-Herrero}}]{RN2158}%
  \BibitemOpen
  \bibfield  {author} {\bibinfo {author} {\bibfnamefont {S.-Y.}\ \bibnamefont
  {Xu}}, \bibinfo {author} {\bibfnamefont {Q.}~\bibnamefont {Ma}}, \bibinfo
  {author} {\bibfnamefont {H.}~\bibnamefont {Shen}}, \bibinfo {author}
  {\bibfnamefont {V.}~\bibnamefont {Fatemi}}, \bibinfo {author} {\bibfnamefont
  {S.}~\bibnamefont {Wu}}, \bibinfo {author} {\bibfnamefont {T.-R.}\
  \bibnamefont {Chang}}, \bibinfo {author} {\bibfnamefont {G.}~\bibnamefont
  {Chang}}, \bibinfo {author} {\bibfnamefont {A.~M.~M.}\ \bibnamefont
  {Valdivia}}, \bibinfo {author} {\bibfnamefont {C.-K.}\ \bibnamefont {Chan}},
  \bibinfo {author} {\bibfnamefont {Q.~D.}\ \bibnamefont {Gibson}}, \bibinfo
  {author} {\bibfnamefont {J.}~\bibnamefont {Zhou}}, \bibinfo {author}
  {\bibfnamefont {Z.}~\bibnamefont {Liu}}, \bibinfo {author} {\bibfnamefont
  {K.}~\bibnamefont {Watanabe}}, \bibinfo {author} {\bibfnamefont
  {T.}~\bibnamefont {Taniguchi}}, \bibinfo {author} {\bibfnamefont
  {H.}~\bibnamefont {Lin}}, \bibinfo {author} {\bibfnamefont {R.~J.}\
  \bibnamefont {Cava}}, \bibinfo {author} {\bibfnamefont {L.}~\bibnamefont
  {Fu}}, \bibinfo {author} {\bibfnamefont {N.}~\bibnamefont {Gedik}}, \ and\
  \bibinfo {author} {\bibfnamefont {P.}~\bibnamefont {Jarillo-Herrero}},\
  }\href {\doibase 10.1038/s41567-018-0189-6} {\bibfield  {journal} {\bibinfo
  {journal} {Nat. Phys.}\ }\textbf {\bibinfo {volume} {14}},\ \bibinfo {pages}
  {900} (\bibinfo {year} {2018})}\BibitemShut {NoStop}%
\bibitem [{\citenamefont {Ashton}\ \emph {et~al.}(2017)\citenamefont {Ashton},
  \citenamefont {Paul}, \citenamefont {Sinnott},\ and\ \citenamefont
  {Hennig}}]{RN1966}%
  \BibitemOpen
  \bibfield  {author} {\bibinfo {author} {\bibfnamefont {M.}~\bibnamefont
  {Ashton}}, \bibinfo {author} {\bibfnamefont {J.}~\bibnamefont {Paul}},
  \bibinfo {author} {\bibfnamefont {S.~B.}\ \bibnamefont {Sinnott}}, \ and\
  \bibinfo {author} {\bibfnamefont {R.~G.}\ \bibnamefont {Hennig}},\ }\href
  {\doibase 10.1103/PhysRevLett.118.106101} {\bibfield  {journal} {\bibinfo
  {journal} {Phys. Rev. Lett.}\ }\textbf {\bibinfo {volume} {118}},\ \bibinfo
  {pages} {106101} (\bibinfo {year} {2017})}\BibitemShut {NoStop}%
\bibitem [{\citenamefont {Mounet}\ \emph {et~al.}(2018)\citenamefont {Mounet},
  \citenamefont {Gibertini}, \citenamefont {Schwaller}, \citenamefont {Campi},
  \citenamefont {Merkys}, \citenamefont {Marrazzo}, \citenamefont {Sohier},
  \citenamefont {Castelli}, \citenamefont {Cepellotti}, \citenamefont {Pizzi},\
  and\ \citenamefont {Marzari}}]{RN1891}%
  \BibitemOpen
  \bibfield  {author} {\bibinfo {author} {\bibfnamefont {N.}~\bibnamefont
  {Mounet}}, \bibinfo {author} {\bibfnamefont {M.}~\bibnamefont {Gibertini}},
  \bibinfo {author} {\bibfnamefont {P.}~\bibnamefont {Schwaller}}, \bibinfo
  {author} {\bibfnamefont {D.}~\bibnamefont {Campi}}, \bibinfo {author}
  {\bibfnamefont {A.}~\bibnamefont {Merkys}}, \bibinfo {author} {\bibfnamefont
  {A.}~\bibnamefont {Marrazzo}}, \bibinfo {author} {\bibfnamefont
  {T.}~\bibnamefont {Sohier}}, \bibinfo {author} {\bibfnamefont {I.~E.}\
  \bibnamefont {Castelli}}, \bibinfo {author} {\bibfnamefont {A.}~\bibnamefont
  {Cepellotti}}, \bibinfo {author} {\bibfnamefont {G.}~\bibnamefont {Pizzi}}, \
  and\ \bibinfo {author} {\bibfnamefont {N.}~\bibnamefont {Marzari}},\ }\href
  {\doibase 10.1038/s41565-017-0035-5} {\bibfield  {journal} {\bibinfo
  {journal} {Nat. Nanotechnol.}\ }\textbf {\bibinfo {volume} {13}},\ \bibinfo
  {pages} {246} (\bibinfo {year} {2018})}\BibitemShut {NoStop}%
\bibitem [{\citenamefont {You}\ \emph {et~al.}(2018)\citenamefont {You},
  \citenamefont {Fang}, \citenamefont {Xu}, \citenamefont {Kaxiras},\ and\
  \citenamefont {Low}}]{RN2113}%
  \BibitemOpen
  \bibfield  {author} {\bibinfo {author} {\bibfnamefont {J.-S.}\ \bibnamefont
  {You}}, \bibinfo {author} {\bibfnamefont {S.}~\bibnamefont {Fang}}, \bibinfo
  {author} {\bibfnamefont {S.-Y.}\ \bibnamefont {Xu}}, \bibinfo {author}
  {\bibfnamefont {E.}~\bibnamefont {Kaxiras}}, \ and\ \bibinfo {author}
  {\bibfnamefont {T.}~\bibnamefont {Low}},\ }\href {\doibase
  10.1103/PhysRevB.98.121109} {\bibfield  {journal} {\bibinfo  {journal} {Phys.
  Rev. B}\ }\textbf {\bibinfo {volume} {98}},\ \bibinfo {pages} {121109}
  (\bibinfo {year} {2018})}\BibitemShut {NoStop}%
\bibitem [{\citenamefont {Zhou}\ \emph {et~al.}(2019)\citenamefont {Zhou},
  \citenamefont {Zhang},\ and\ \citenamefont {Law}}]{RN2172}%
  \BibitemOpen
  \bibfield  {author} {\bibinfo {author} {\bibfnamefont {B.~T.}\ \bibnamefont
  {Zhou}}, \bibinfo {author} {\bibfnamefont {C.-P.}\ \bibnamefont {Zhang}}, \
  and\ \bibinfo {author} {\bibfnamefont {K.~T.}\ \bibnamefont {Law}},\
  }\href@noop {} {\bibfield  {journal} {\bibinfo  {journal} {arXiv:1903.11958}\
  } (\bibinfo {year} {2019})}\BibitemShut {NoStop}%
\bibitem [{\citenamefont {Giannozzi}\ \emph {et~al.}(2009)\citenamefont
  {Giannozzi}, \citenamefont {Baroni}, \citenamefont {Bonini}, \citenamefont
  {Calandra}, \citenamefont {Car}, \citenamefont {Cavazzoni}, \citenamefont
  {Ceresoli}, \citenamefont {Chiarotti}, \citenamefont {Cococcioni},
  \citenamefont {Dabo}, \citenamefont {Corso}, \citenamefont {Gironcoli},
  \citenamefont {Fabris}, \citenamefont {Fratesi}, \citenamefont {Gebauer},
  \citenamefont {Gerstmann}, \citenamefont {Gougoussis}, \citenamefont
  {Kokalj}, \citenamefont {Lazzeri}, \citenamefont {Martin-Samos},
  \citenamefont {Marzari}, \citenamefont {Mauri}, \citenamefont {Mazzarello},
  \citenamefont {Paolini}, \citenamefont {Pasquarello}, \citenamefont
  {Paulatto}, \citenamefont {Sbraccia}, \citenamefont {Scandolo}, \citenamefont
  {Sclauzero}, \citenamefont {Seitsonen}, \citenamefont {Smogunov},
  \citenamefont {Umari},\ and\ \citenamefont {Wentzcovitch}}]{RN82}%
  \BibitemOpen
  \bibfield  {author} {\bibinfo {author} {\bibfnamefont {P.}~\bibnamefont
  {Giannozzi}}, \bibinfo {author} {\bibfnamefont {S.}~\bibnamefont {Baroni}},
  \bibinfo {author} {\bibfnamefont {N.}~\bibnamefont {Bonini}}, \bibinfo
  {author} {\bibfnamefont {M.}~\bibnamefont {Calandra}}, \bibinfo {author}
  {\bibfnamefont {R.}~\bibnamefont {Car}}, \bibinfo {author} {\bibfnamefont
  {C.}~\bibnamefont {Cavazzoni}}, \bibinfo {author} {\bibfnamefont
  {D.}~\bibnamefont {Ceresoli}}, \bibinfo {author} {\bibfnamefont {G.~L.}\
  \bibnamefont {Chiarotti}}, \bibinfo {author} {\bibfnamefont {M.}~\bibnamefont
  {Cococcioni}}, \bibinfo {author} {\bibfnamefont {I.}~\bibnamefont {Dabo}},
  \bibinfo {author} {\bibfnamefont {A.~D.}\ \bibnamefont {Corso}}, \bibinfo
  {author} {\bibfnamefont {S.~d.}\ \bibnamefont {Gironcoli}}, \bibinfo {author}
  {\bibfnamefont {S.}~\bibnamefont {Fabris}}, \bibinfo {author} {\bibfnamefont
  {G.}~\bibnamefont {Fratesi}}, \bibinfo {author} {\bibfnamefont
  {R.}~\bibnamefont {Gebauer}}, \bibinfo {author} {\bibfnamefont
  {U.}~\bibnamefont {Gerstmann}}, \bibinfo {author} {\bibfnamefont
  {C.}~\bibnamefont {Gougoussis}}, \bibinfo {author} {\bibfnamefont
  {A.}~\bibnamefont {Kokalj}}, \bibinfo {author} {\bibfnamefont
  {M.}~\bibnamefont {Lazzeri}}, \bibinfo {author} {\bibfnamefont
  {L.}~\bibnamefont {Martin-Samos}}, \bibinfo {author} {\bibfnamefont
  {N.}~\bibnamefont {Marzari}}, \bibinfo {author} {\bibfnamefont
  {F.}~\bibnamefont {Mauri}}, \bibinfo {author} {\bibfnamefont
  {R.}~\bibnamefont {Mazzarello}}, \bibinfo {author} {\bibfnamefont
  {S.}~\bibnamefont {Paolini}}, \bibinfo {author} {\bibfnamefont
  {A.}~\bibnamefont {Pasquarello}}, \bibinfo {author} {\bibfnamefont
  {L.}~\bibnamefont {Paulatto}}, \bibinfo {author} {\bibfnamefont
  {C.}~\bibnamefont {Sbraccia}}, \bibinfo {author} {\bibfnamefont
  {S.}~\bibnamefont {Scandolo}}, \bibinfo {author} {\bibfnamefont
  {G.}~\bibnamefont {Sclauzero}}, \bibinfo {author} {\bibfnamefont {A.~P.}\
  \bibnamefont {Seitsonen}}, \bibinfo {author} {\bibfnamefont {A.}~\bibnamefont
  {Smogunov}}, \bibinfo {author} {\bibfnamefont {P.}~\bibnamefont {Umari}}, \
  and\ \bibinfo {author} {\bibfnamefont {R.~M.}\ \bibnamefont {Wentzcovitch}},\
  }\href {http://stacks.iop.org/0953-8984/21/i=39/a=395502} {\bibfield
  {journal} {\bibinfo  {journal} {J. Phys.: Condens. Matter}\ }\textbf
  {\bibinfo {volume} {21}},\ \bibinfo {pages} {395502} (\bibinfo {year}
  {2009})}\BibitemShut {NoStop}%
\bibitem [{\citenamefont {Mostofi}\ \emph {et~al.}(2008)\citenamefont
  {Mostofi}, \citenamefont {Yates}, \citenamefont {Lee}, \citenamefont {Souza},
  \citenamefont {Vanderbilt},\ and\ \citenamefont {Marzari}}]{RN149}%
  \BibitemOpen
  \bibfield  {author} {\bibinfo {author} {\bibfnamefont {A.~A.}\ \bibnamefont
  {Mostofi}}, \bibinfo {author} {\bibfnamefont {J.~R.}\ \bibnamefont {Yates}},
  \bibinfo {author} {\bibfnamefont {Y.-S.}\ \bibnamefont {Lee}}, \bibinfo
  {author} {\bibfnamefont {I.}~\bibnamefont {Souza}}, \bibinfo {author}
  {\bibfnamefont {D.}~\bibnamefont {Vanderbilt}}, \ and\ \bibinfo {author}
  {\bibfnamefont {N.}~\bibnamefont {Marzari}},\ }\href {\doibase
  10.1016/j.cpc.2007.11.016} {\bibfield  {journal} {\bibinfo  {journal}
  {Comput. Phys. Commun.}\ }\textbf {\bibinfo {volume} {178}},\ \bibinfo
  {pages} {685} (\bibinfo {year} {2008})}\BibitemShut {NoStop}%
\bibitem [{\citenamefont {Mostofi}\ \emph {et~al.}(2014)\citenamefont
  {Mostofi}, \citenamefont {Yates}, \citenamefont {Pizzi}, \citenamefont {Lee},
  \citenamefont {Souza}, \citenamefont {Vanderbilt},\ and\ \citenamefont
  {Marzari}}]{RN772}%
  \BibitemOpen
  \bibfield  {author} {\bibinfo {author} {\bibfnamefont {A.~A.}\ \bibnamefont
  {Mostofi}}, \bibinfo {author} {\bibfnamefont {J.~R.}\ \bibnamefont {Yates}},
  \bibinfo {author} {\bibfnamefont {G.}~\bibnamefont {Pizzi}}, \bibinfo
  {author} {\bibfnamefont {Y.-S.}\ \bibnamefont {Lee}}, \bibinfo {author}
  {\bibfnamefont {I.}~\bibnamefont {Souza}}, \bibinfo {author} {\bibfnamefont
  {D.}~\bibnamefont {Vanderbilt}}, \ and\ \bibinfo {author} {\bibfnamefont
  {N.}~\bibnamefont {Marzari}},\ }\href {\doibase 10.1016/j.cpc.2014.05.003}
  {\bibfield  {journal} {\bibinfo  {journal} {Comput. Phys. Commun.}\ }\textbf
  {\bibinfo {volume} {185}},\ \bibinfo {pages} {2309} (\bibinfo {year}
  {2014})}\BibitemShut {NoStop}%
\bibitem [{\citenamefont {Wu}\ \emph {et~al.}(2018)\citenamefont {Wu},
  \citenamefont {Zhang}, \citenamefont {Song}, \citenamefont {Troyer},\ and\
  \citenamefont {Soluyanov}}]{RN1186}%
  \BibitemOpen
  \bibfield  {author} {\bibinfo {author} {\bibfnamefont {Q.}~\bibnamefont
  {Wu}}, \bibinfo {author} {\bibfnamefont {S.}~\bibnamefont {Zhang}}, \bibinfo
  {author} {\bibfnamefont {H.-F.}\ \bibnamefont {Song}}, \bibinfo {author}
  {\bibfnamefont {M.}~\bibnamefont {Troyer}}, \ and\ \bibinfo {author}
  {\bibfnamefont {A.~A.}\ \bibnamefont {Soluyanov}},\ }\href {\doibase
  doi.org/10.1016/j.cpc.2017.09.033} {\bibfield  {journal} {\bibinfo  {journal}
  {Comput. Phys. Commun.}\ }\textbf {\bibinfo {volume} {224}},\ \bibinfo
  {pages} {405} (\bibinfo {year} {2018})}\BibitemShut {NoStop}%
\bibitem [{\citenamefont {Ryu}\ \emph {et~al.}(2018)\citenamefont {Ryu},
  \citenamefont {Chen}, \citenamefont {Kim}, \citenamefont {Tsai},
  \citenamefont {Tang}, \citenamefont {Jiang}, \citenamefont {Liou},
  \citenamefont {Kahn}, \citenamefont {Jia}, \citenamefont {Omrani},
  \citenamefont {Shim}, \citenamefont {Hussain}, \citenamefont {Shen},
  \citenamefont {Kim}, \citenamefont {Min}, \citenamefont {Hwang},
  \citenamefont {Crommie},\ and\ \citenamefont {Mo}}]{RN2204}%
  \BibitemOpen
  \bibfield  {author} {\bibinfo {author} {\bibfnamefont {H.}~\bibnamefont
  {Ryu}}, \bibinfo {author} {\bibfnamefont {Y.}~\bibnamefont {Chen}}, \bibinfo
  {author} {\bibfnamefont {H.}~\bibnamefont {Kim}}, \bibinfo {author}
  {\bibfnamefont {H.~Z.}\ \bibnamefont {Tsai}}, \bibinfo {author}
  {\bibfnamefont {S.}~\bibnamefont {Tang}}, \bibinfo {author} {\bibfnamefont
  {J.}~\bibnamefont {Jiang}}, \bibinfo {author} {\bibfnamefont
  {F.}~\bibnamefont {Liou}}, \bibinfo {author} {\bibfnamefont {S.}~\bibnamefont
  {Kahn}}, \bibinfo {author} {\bibfnamefont {C.}~\bibnamefont {Jia}}, \bibinfo
  {author} {\bibfnamefont {A.~A.}\ \bibnamefont {Omrani}}, \bibinfo {author}
  {\bibfnamefont {J.~H.}\ \bibnamefont {Shim}}, \bibinfo {author}
  {\bibfnamefont {Z.}~\bibnamefont {Hussain}}, \bibinfo {author} {\bibfnamefont
  {Z.~X.}\ \bibnamefont {Shen}}, \bibinfo {author} {\bibfnamefont
  {K.}~\bibnamefont {Kim}}, \bibinfo {author} {\bibfnamefont {B.~I.}\
  \bibnamefont {Min}}, \bibinfo {author} {\bibfnamefont {C.}~\bibnamefont
  {Hwang}}, \bibinfo {author} {\bibfnamefont {M.~F.}\ \bibnamefont {Crommie}},
  \ and\ \bibinfo {author} {\bibfnamefont {S.~K.}\ \bibnamefont {Mo}},\ }\href
  {\doibase 10.1021/acs.nanolett.7b03264} {\bibfield  {journal} {\bibinfo
  {journal} {Nano Lett.}\ }\textbf {\bibinfo {volume} {18}},\ \bibinfo {pages}
  {689} (\bibinfo {year} {2018})}\BibitemShut {NoStop}%
\bibitem [{\citenamefont {Staley}\ \emph {et~al.}(2009)\citenamefont {Staley},
  \citenamefont {Wu}, \citenamefont {Eklund}, \citenamefont {Liu},
  \citenamefont {Li},\ and\ \citenamefont {Xu}}]{RN1489}%
  \BibitemOpen
  \bibfield  {author} {\bibinfo {author} {\bibfnamefont {N.~E.}\ \bibnamefont
  {Staley}}, \bibinfo {author} {\bibfnamefont {J.}~\bibnamefont {Wu}}, \bibinfo
  {author} {\bibfnamefont {P.}~\bibnamefont {Eklund}}, \bibinfo {author}
  {\bibfnamefont {Y.}~\bibnamefont {Liu}}, \bibinfo {author} {\bibfnamefont
  {L.}~\bibnamefont {Li}}, \ and\ \bibinfo {author} {\bibfnamefont
  {Z.}~\bibnamefont {Xu}},\ }\href
  {https://link.aps.org/doi/10.1103/PhysRevB.80.184505} {\bibfield  {journal}
  {\bibinfo  {journal} {Phys. Rev. B}\ }\textbf {\bibinfo {volume} {80}},\
  \bibinfo {pages} {184505} (\bibinfo {year} {2009})}\BibitemShut {NoStop}%
\bibitem [{\citenamefont {Ugeda}\ \emph {et~al.}(2016)\citenamefont {Ugeda},
  \citenamefont {Bradley}, \citenamefont {Zhang}, \citenamefont {Onishi},
  \citenamefont {Chen}, \citenamefont {Ruan}, \citenamefont
  {Ojeda-Aristizabal}, \citenamefont {Ryu}, \citenamefont {Edmonds},
  \citenamefont {Tsai}, \citenamefont {Riss}, \citenamefont {Mo}, \citenamefont
  {Lee}, \citenamefont {Zettl}, \citenamefont {Hussain}, \citenamefont {Shen},\
  and\ \citenamefont {Crommie}}]{RN1470}%
  \BibitemOpen
  \bibfield  {author} {\bibinfo {author} {\bibfnamefont {M.~M.}\ \bibnamefont
  {Ugeda}}, \bibinfo {author} {\bibfnamefont {A.~J.}\ \bibnamefont {Bradley}},
  \bibinfo {author} {\bibfnamefont {Y.}~\bibnamefont {Zhang}}, \bibinfo
  {author} {\bibfnamefont {S.}~\bibnamefont {Onishi}}, \bibinfo {author}
  {\bibfnamefont {Y.}~\bibnamefont {Chen}}, \bibinfo {author} {\bibfnamefont
  {W.}~\bibnamefont {Ruan}}, \bibinfo {author} {\bibfnamefont {C.}~\bibnamefont
  {Ojeda-Aristizabal}}, \bibinfo {author} {\bibfnamefont {H.}~\bibnamefont
  {Ryu}}, \bibinfo {author} {\bibfnamefont {M.~T.}\ \bibnamefont {Edmonds}},
  \bibinfo {author} {\bibfnamefont {H.-Z.}\ \bibnamefont {Tsai}}, \bibinfo
  {author} {\bibfnamefont {A.}~\bibnamefont {Riss}}, \bibinfo {author}
  {\bibfnamefont {S.-K.}\ \bibnamefont {Mo}}, \bibinfo {author} {\bibfnamefont
  {D.}~\bibnamefont {Lee}}, \bibinfo {author} {\bibfnamefont {A.}~\bibnamefont
  {Zettl}}, \bibinfo {author} {\bibfnamefont {Z.}~\bibnamefont {Hussain}},
  \bibinfo {author} {\bibfnamefont {Z.-X.}\ \bibnamefont {Shen}}, \ and\
  \bibinfo {author} {\bibfnamefont {M.~F.}\ \bibnamefont {Crommie}},\ }\href
  {\doibase 10.1038/nphys3527
  http://www.nature.com/nphys/journal/v12/n1/abs/nphys3527.html#supplementary-information}
  {\bibfield  {journal} {\bibinfo  {journal} {Nat. Phys.}\ }\textbf {\bibinfo
  {volume} {12}},\ \bibinfo {pages} {92} (\bibinfo {year} {2016})}\BibitemShut
  {NoStop}%
\bibitem [{\citenamefont {Xi}\ \emph {et~al.}(2016{\natexlab{a}})\citenamefont
  {Xi}, \citenamefont {Berger}, \citenamefont {Forr車}, \citenamefont {Shan},\
  and\ \citenamefont {Mak}}]{RN1540}%
  \BibitemOpen
  \bibfield  {author} {\bibinfo {author} {\bibfnamefont {X.}~\bibnamefont
  {Xi}}, \bibinfo {author} {\bibfnamefont {H.}~\bibnamefont {Berger}}, \bibinfo
  {author} {\bibfnamefont {L.}~\bibnamefont {Forr車}}, \bibinfo {author}
  {\bibfnamefont {J.}~\bibnamefont {Shan}}, \ and\ \bibinfo {author}
  {\bibfnamefont {K.~F.}\ \bibnamefont {Mak}},\ }\href
  {https://link.aps.org/doi/10.1103/PhysRevLett.117.106801} {\bibfield
  {journal} {\bibinfo  {journal} {Phys. Rev. Lett.}\ }\textbf {\bibinfo
  {volume} {117}},\ \bibinfo {pages} {106801} (\bibinfo {year}
  {2016}{\natexlab{a}})}\BibitemShut {NoStop}%
\bibitem [{\citenamefont {Xi}\ \emph {et~al.}(2016{\natexlab{b}})\citenamefont
  {Xi}, \citenamefont {Wang}, \citenamefont {Zhao}, \citenamefont {Park},
  \citenamefont {Law}, \citenamefont {Berger}, \citenamefont {Forro},
  \citenamefont {Shan},\ and\ \citenamefont {Mak}}]{RN1521}%
  \BibitemOpen
  \bibfield  {author} {\bibinfo {author} {\bibfnamefont {X.}~\bibnamefont
  {Xi}}, \bibinfo {author} {\bibfnamefont {Z.}~\bibnamefont {Wang}}, \bibinfo
  {author} {\bibfnamefont {W.}~\bibnamefont {Zhao}}, \bibinfo {author}
  {\bibfnamefont {J.-H.}\ \bibnamefont {Park}}, \bibinfo {author}
  {\bibfnamefont {K.~T.}\ \bibnamefont {Law}}, \bibinfo {author} {\bibfnamefont
  {H.}~\bibnamefont {Berger}}, \bibinfo {author} {\bibfnamefont
  {L.}~\bibnamefont {Forro}}, \bibinfo {author} {\bibfnamefont
  {J.}~\bibnamefont {Shan}}, \ and\ \bibinfo {author} {\bibfnamefont {K.~F.}\
  \bibnamefont {Mak}},\ }\href {\doibase 10.1038/nphys3538
  http://www.nature.com/nphys/journal/v12/n2/abs/nphys3538.html#supplementary-information}
  {\bibfield  {journal} {\bibinfo  {journal} {Nat. Phys.}\ }\textbf {\bibinfo
  {volume} {12}},\ \bibinfo {pages} {139} (\bibinfo {year}
  {2016}{\natexlab{b}})}\BibitemShut {NoStop}%
\bibitem [{\citenamefont {Xi}\ \emph {et~al.}(2015)\citenamefont {Xi},
  \citenamefont {Zhao}, \citenamefont {Wang}, \citenamefont {Berger},
  \citenamefont {Forr車}, \citenamefont {Shan},\ and\ \citenamefont
  {Mak}}]{RN1522}%
  \BibitemOpen
  \bibfield  {author} {\bibinfo {author} {\bibfnamefont {X.}~\bibnamefont
  {Xi}}, \bibinfo {author} {\bibfnamefont {L.}~\bibnamefont {Zhao}}, \bibinfo
  {author} {\bibfnamefont {Z.}~\bibnamefont {Wang}}, \bibinfo {author}
  {\bibfnamefont {H.}~\bibnamefont {Berger}}, \bibinfo {author} {\bibfnamefont
  {L.}~\bibnamefont {Forr車}}, \bibinfo {author} {\bibfnamefont
  {J.}~\bibnamefont {Shan}}, \ and\ \bibinfo {author} {\bibfnamefont {K.~F.}\
  \bibnamefont {Mak}},\ }\href {\doibase 10.1038/nnano.2015.143} {\bibfield
  {journal} {\bibinfo  {journal} {Nat. Nano.}\ }\textbf {\bibinfo {volume}
  {10}},\ \bibinfo {pages} {765} (\bibinfo {year} {2015})}\BibitemShut
  {NoStop}%
\bibitem [{\citenamefont {Son}\ \emph {et~al.}(2019)\citenamefont {Son},
  \citenamefont {Kim}, \citenamefont {Ahn}, \citenamefont {Lee},\ and\
  \citenamefont {Lee}}]{RN2243}%
  \BibitemOpen
  \bibfield  {author} {\bibinfo {author} {\bibfnamefont {J.}~\bibnamefont
  {Son}}, \bibinfo {author} {\bibfnamefont {K.-H.}\ \bibnamefont {Kim}},
  \bibinfo {author} {\bibfnamefont {Y.~H.}\ \bibnamefont {Ahn}}, \bibinfo
  {author} {\bibfnamefont {H.-W.}\ \bibnamefont {Lee}}, \ and\ \bibinfo
  {author} {\bibfnamefont {J.}~\bibnamefont {Lee}},\ }\href@noop {} {\bibfield
  {journal} {\bibinfo  {journal} {arXiv:1907.00010}\ } (\bibinfo {year}
  {2019})}\BibitemShut {NoStop}%
\bibitem [{\citenamefont {Du}\ \emph {et~al.}(2018)\citenamefont {Du},
  \citenamefont {Wang}, \citenamefont {Lu},\ and\ \citenamefont
  {Xie}}]{RN2109}%
  \BibitemOpen
  \bibfield  {author} {\bibinfo {author} {\bibfnamefont {Z.~Z.}\ \bibnamefont
  {Du}}, \bibinfo {author} {\bibfnamefont {C.~M.}\ \bibnamefont {Wang}},
  \bibinfo {author} {\bibfnamefont {H.~Z.}\ \bibnamefont {Lu}}, \ and\ \bibinfo
  {author} {\bibfnamefont {X.~C.}\ \bibnamefont {Xie}},\ }\href {\doibase
  10.1103/PhysRevLett.121.266601} {\bibfield  {journal} {\bibinfo  {journal}
  {Phys. Rev. Lett.}\ }\textbf {\bibinfo {volume} {121}},\ \bibinfo {pages}
  {266601} (\bibinfo {year} {2018})}\BibitemShut {NoStop}%
\bibitem [{\citenamefont {Naito}\ and\ \citenamefont {Tanaka}(1982)}]{RN2253}%
  \BibitemOpen
  \bibfield  {author} {\bibinfo {author} {\bibfnamefont {M.}~\bibnamefont
  {Naito}}\ and\ \bibinfo {author} {\bibfnamefont {S.}~\bibnamefont {Tanaka}},\
  }\href {\doibase 10.1143/JPSJ.51.219} {\bibfield  {journal} {\bibinfo
  {journal} {J. Phys. Soc. Jpn.}\ }\textbf {\bibinfo {volume} {51}},\ \bibinfo
  {pages} {219} (\bibinfo {year} {1982})}\BibitemShut {NoStop}%
\bibitem [{\citenamefont {Zhang}\ \emph {et~al.}(2018)\citenamefont {Zhang},
  \citenamefont {van~den Brink}, \citenamefont {Felser},\ and\ \citenamefont
  {Yan}}]{RN2112}%
  \BibitemOpen
  \bibfield  {author} {\bibinfo {author} {\bibfnamefont {Y.}~\bibnamefont
  {Zhang}}, \bibinfo {author} {\bibfnamefont {J.}~\bibnamefont {van~den
  Brink}}, \bibinfo {author} {\bibfnamefont {C.}~\bibnamefont {Felser}}, \ and\
  \bibinfo {author} {\bibfnamefont {B.}~\bibnamefont {Yan}},\ }\href {\doibase
  10.1088/2053-1583/aad1ae} {\bibfield  {journal} {\bibinfo  {journal} {2D
  Mater.}\ }\textbf {\bibinfo {volume} {5}},\ \bibinfo {pages} {044001}
  (\bibinfo {year} {2018})}\BibitemShut {NoStop}%
\end{thebibliography}
%

\end {document}